\documentstyle[preprint,prb,aps]{revtex}

\begin{document}

%\tighten

\draft

\preprint{Submitted to Physical Review B}

\author{Neeraj Mainkar, D. A. Browne and J. Callaway}

\address{
Department of Physics and Astronomy\\
Louisiana State University\\
Baton Rouge, Louisiana 70803}

\date{\today}

\title{ First Principles LCGO calculation of the magneto-optical properties
of nickel and iron}

\maketitle

\begin{abstract}
We report a first principles, self-consistent, all electron, linear
combination of Gaussian orbitals (LCGO) calculation of a comprehensive
collection of magneto-optical properties of nickel and iron based on
density functional theory.  Among the many magneto-optical effects, we
have studied the equatorial Kerr effect for absorption in the optical
as well as soft X-ray region, where it is called X-ray magnetic linear
dichroism (X-MLD). In the optical region the effect is  of the order
of 2\% while in the X-ray region it is of the order of 1\% for the
incident angles considered.  In addition, the polar Kerr effect, X-ray
magnetic circular dichroism (X-MCD) and total X-ray absorption at the
L$_{2,3}$ edges, soft X-ray Faraday effect at the L$_{2,3}$ edges have also
been calculated.  Our results are in good agreement with experiments and
other first principles methods that have been used to calculate some of
these properties.

\end{abstract}


\narrowtext

\section{INTRODUCTION}

Although magneto-optical properties of magnetic metals have been known
for over a hundred years,\cite{Voigt} it is only in the past couple of
decades that vigorous interest has been refocussed on this subject,
partly due to their potential for application in the technology of high
density data storage.\cite{Schlic,Stohr,Meikle,Kryder} Because of
advances in laser and tunable synchrotron sources, a variety of
different magneto-optical effects including the magneto-optical Kerr
effect (MOKE),\cite{Zhai} the Faraday effect,\cite{Kortright} X-ray
magnetic circular dichroism for absorption\cite{Schutz,Chen} as well as
angle resolved photoemission,\cite{Baumgar,Laan} and X-magnetic linear
dichroism in angle resolved photoemission\cite{Laan,Henk} have now been
extensively studied.  The connection between all of these phenomena and
the electronic structure of the materials in which they are seen, has
long been known.\cite{Bennet,Erskine} However, only in recent years has
there been an effort to perform first principles band structure
calculations of some of these effects.  One electron band calculations
for bulk and surfaces, alloys and multilayers have been already
reported for the polar MOKE,\cite{Oppeneer1,Oppeneer2,Maurer,Guo1,Guo2}
X-MCD and X-MLD,\cite{Ebert1,Ebert2,Guo3,Wu1,Wu2,Igarashi} and the
X-ray Faraday effect.\cite{Gotsis}

This paper reports a first principles, self-consistent, all electron
calculation of several magneto-optical properties of bulk nickel and
iron, using the linear combination of Gaussian orbitals (LCGO) method.
These include the equatorial Kerr effect, X-ray magnetic linear
dichroism (X-MLD) in absorption at the $2p$ edges, the soft X-ray
Faraday effect at the $2p$ edges, the X-MCD and absorption spectra at
the $2p$ edges as well as the polar MOKE.  The equatorial Kerr effect
defined here is the difference in absorption of $p$-polarized light
incident obliquely on the magnetized sample in the equatorial geometry,
when the sample magnetization is reversed.  When observed at core-level
edges this effect can be called photoabsorption X-MLD of a new kind to
distinguish it from the first kind observed by holding the photon
polarization parallel to the $x$-axis and rotating the magnetization
from the $x$ to the $z$ axis,\cite{Guo4} the second kind obseved due to
the rotation of the photon polarization vector with the magnetization
held fixed\cite{Guo3} and the one observed in angle resolved
photemission.

Our motivation in performing these calculations has been two-fold.
Firstly, since it is known that these effects depend sensitively on the
accuracy of the band structure calculation, we wanted to demonstrate
that our modified tight binding method that includes spin-orbit
coupling in a very straightforward manner, can produce results at least
as accurate as other {\em ab-initio} electronic structure approaches
that apply either the fully relativistic machinery or use more
complicated computational procedures to avoid using the Kramers Kronig
(KK) transformation.  Our MOKE results on nickel and iron essentially
prove this point.  Secondly, although experiments on the equatorial
Kerr effect have been done previously in the optical
region,\cite{Krigorb} no experiments on the above-mentioned
photoabsorption X-MLD at the $2p$ edge of nickel and iron have been
reported.  It would thus be interesting to see how well our theoretical
results for the $2p$ edge agree with future experiments.  A recent
experiment on the soft X-ray Faraday effect at the $2p$ edges of
iron\cite{Kortright} has also spurred us into calculating this effect
using the LCGO method.

We thus begin Section II by outlining in brief the inclusion of
spin-orbit coupling to the non-relativistic LCGO method, which is
slightly different from the previous LCGO work.\cite{WangCal}  We also
give a brief description of the fast and efficient KK transformation
method we have employed in our analysis.  In section III we discuss in
detail the results of our first principles calculation of the elements
of the conductivity tensor and their subsequent use in determining the
magneto-optical properties mentioned above.  We shall also point out
how our method compares with previous theoretical results and with
available data.

\section{Method of Calculation}

The band structure calculation with spin-orbit coupling included within
the LCGO method was first done by Wang and Callaway\cite{WangCal} for
nickel and by Singh, Wang and Callaway for iron.\cite{Singh}  However,
the exchange potential in those calculations was of the X$\alpha$ kind
and the set of 38 basis functions was also small.  Subsequently, the
computational procedure was revised considerably,\cite{Wang} with the
exchange potential being replaced by the more accurate von-Barth and
Hedin type as parametrized by Rajagopal, Singhal and
Kimball,\cite{Raja} and the basis set expanded to a total of 75
functions for $3d$ transition metals.  The method of evaluating the
Brillouin zone (BZ) integrals was also improved by using the linear
analytic tetrahedron method, \cite{Rath,Singhal} and improvements were
made in calculating the $K = 0$ Fourier component of the Coulomb
potential.\cite{Blaha} This non-relativistic LCGO method was applied to
a host of different elemental magnetic as well as non-magnetic solids
over the past several years and a variety of electronic properties such
as the Fermi surface, the optical conductivity and the Compton profile
were calculated
\cite{Callaway,Wang2,Laurent1,Laurent2,Laurent3,Jani1,Trip,Hchen,Jani2}
which were in fairly good agreement with experiments.  For the 3d
transition metals the basis set thus consisted of 13 s-type, 10 p-type,
5 d-type and 1 f-type Gaussian orbitals, using published orbitals based
on atomic calculations.\cite{Wachters}  However, relativistic
corrections were completely ignored in these improvements.

With a view to tackling the anisotropic properties of magnetic metals,
we have added the spin-orbit coupling, Darwin and relativistic
mass-velocity terms to the existing code.  As is well-known, the
principal advantage to using a Gaussian basis in a first principles
tight-binding method is the ability to use analytic expressions for the
overlap and several Hamiltonian matrix elements.  This is true in the
case of the spin orbit interaction term as well.  We have used
essentially the same approximations used earlier\cite{WangCal} for the
spin-orbit coupling term {\em i.e.}, the principal contribution to the
spin-orbit matrix elements is only for those in the $p-p$ and $d-d$
blocks.  We also used the same central-cell approximation wherein we
retain only spin-orbit matrix elements between those orbitals centered
on the same atomic site.  The difference between our new approach and
theirs is that in Ref.~\protect\onlinecite{WangCal} the spin orbit
matrix elements were evaluated in reciprocal space and a Fourier sum
had to be performed over $K$ vectors to obtain the real space
potential.  Since the spin-orbit coupling is strongest at smaller
distances, this necessitated the use of an asymptotic expansion for the
handling of larger $K$ values to obtain a reasonable value for the $K$
summation.  We have avoided this completely by evaluating the
spin-orbit matrix elements in real space directly.  In terms of the
Gaussian basis, the expresssion for the spin orbit matrix elements of
the $p-p$ block is,

\begin{equation}
I_p = {\frac{\hbar^{2} e^{2} N_{p}}{8 m^{2}c^{2}\alpha}}
\biggl[ Z - {\frac{1}{2 N \Omega}}\sum_{i,j}^{n_b}\frac{P_{ij} N_{ij}\Gamma(
(l_{ij} + 3)/2)}{(\alpha + \alpha_{ij})^{\frac{l_{ij} +3}{2}}} \biggr]
\label{eq:sop}
\end{equation}
while for those of the $d-d$ block is
\begin{equation}
I_d = {\frac{\hbar^{2} e^{2} N_{d}}{8 m^{2}c^{2}\alpha^{2}}}
\biggl[ Z - {\frac{1}{2 N \Omega}}\sum_{i,j}^{n_b}P_{ij}
N_{ij}\biggl\{\frac{\Gamma((l_{ij} + 3)/2)}{(\alpha +
\alpha_{ij})^{\frac{l_{ij} +3}{2}}} + {\frac{\alpha\Gamma((l_{ij} +
5)/2)}{(\alpha + \alpha_{ij})^{\frac{l_{ij} + 5}{2}}}}\biggl\}\biggl]
\label{eq:sod}
\end{equation}
where
\begin{equation}
P_{ij} = {\frac{1}{48W}}\sum_{k}\rho_{ij}^{occ}(k)g(k)
\label{eq:Pij}
\end{equation}

In the above expressions $N_{p}$ and $N_{d}$ are the product of the
normalization constants of the appropriate Gaussian orbitals for the
$p-p$ and $d-d$ block respectively, $n_b$ is the number of basis
functions, $\alpha$ is the sum of the exponents of the Gaussian
orbitals, $N$ is the number of unit cells and $\Omega$ is the unit cell
volume, $g(k)$ is the weight associated with each $k$ point in the BZ,
$l_{ij}$ is the sum of the orbital quantum numbers of the basis
functions $i$ and $j$, $W = \sum_{k}g(k)$ is the total weight, and
$\Gamma(z)$ is the Gamma function.  For the charge density appearing in
the Eq.~(\ref{eq:Pij}) we used the resulting wavefunctions of the
non-relativistic, self-consistent calculation.  With spin-orbit
coupling included, the new generalized eigenvalue equation of order
150$\times$150 was diagonalized in $1/16$th of the BZ at only $219$ $k$
points for fcc nickel and only $125$ $k$ points for bcc iron, fewer
than the number of points used in the previous calculations.

Using the results of the band structure calculation, the elements of
the conductivity tensor may be found from the formulas\cite{WangCal},
\begin{equation}
\sigma_{xx}\/(\omega) = \frac{ie^{2}}{m^{2}\hbar}\sum_k
\sum_{ln} \frac{1}{\omega_{nl}(k)}
\biggl[\frac{|\Pi_{ln}^{x}|^2} {\omega - \omega_{nl}(k) +i\delta}
+ \frac{|\Pi_{ln}^{x}|^2}{\omega +\omega_{nl}(k) +i\delta} \biggr]
\label{eq:kubo1}
\end{equation}
\begin{equation}
\sigma_{xy}\/(\omega) = \frac{ie^{2}}{m^{2}\hbar}\sum_k
\sum_{ln} \frac{1}{\omega_{nl}(k)} \biggl[\frac{\Pi_{ln}^{x}\Pi_{nl}^{y}}
{\omega - \omega_{nl}(k) +i\delta} +
\frac{(\Pi_{ln}^{x}\Pi_{nl}^{y})^{\star}}
{\omega + \omega_{nl}(k) +i\delta} \biggr]
\label{eq:kubo2}
\end{equation}
where $l$ goes over the occupied states and $n$ goes over the unoccupied
states and $\Pi$'s are the $k$ dependent matrix elements of the momentum
operator.

The usual procedure is to evaluate the real part of $\sigma_{xx}$ and
the imagninary part of $\sigma_{xy}$ by replacing the Lorentzian by a
delta function in the limit of $\delta$ going to zero. It is then
customary to keep $\delta$ finite to simulate a finite relaxation time.
For calculations of magneto-optical properties, it is then necessary to
perform the KK transformations to obtain the imaginary part of
$\sigma_{xx}$ and the real part of $\sigma_{xy}$.\cite{Bennet}  The KK
integrals are known to suffer from problems of slow convergence and the
necessity of choosing high cutoff values for energy.  One
approach\cite{Oppeneer1} evaluates the original Kubo formula directly
with lifetime effects as a parameter.  This method, although accurate,
is extremely computationally intensive.  We have instead performed the
KK transformations through the use of two succesive Fast Fourier
transforms.  This method is commonly used in studies of infrared
intensities of liquids.\cite{Bertie}  The basis of this method comes
from the well known relation that if $F(t)$ is the Fourier transform of
$f(\omega)$,
\begin{equation}
F(t) = \frac{1}{\sqrt{2\pi}}\int_{-\infty}^{\infty} d\omega~
f(\omega)~e^{-i{\omega}t},
\label{eq:KK1}
\end{equation}
and $h(\omega)$ is the Hilbert transform of $f(\omega)$,
\begin{equation}
h(\omega) =
\frac{1}{\pi}P\int_{-\infty}^{\infty} d\omega^{\prime}\,
\frac{f(\omega^{\prime})}
{\omega - \omega^{\prime}}
\label{eq:KK2}
\end{equation}
then
\begin{equation}
H(t) = -i~{\rm sgn(t)}~F(t)
\label{eq:KK}
\end{equation}
where $H(t)$ is the Fourier transform of $h(\omega)$,
\begin{equation}
H(t) = \frac{1}{\sqrt{2\pi}}\int_{-\infty}^{\infty}
d\omega\,h(\omega)\,e^{-i{\omega}t}
\label{eq:KK3}
\end{equation}

Preliminary tests on known Hilbert transforms using this method yield
results that are accurate to better than 1\%.  The advantages of this
method are immense since with sufficient number of points, the method
is very accurate and fast.  Furthermore, it has the flexibility of
incorporating the lifetime effects very conveniently by simply
multiplying the right hand side of Eq.~(\ref{eq:KK}) by an exponential
damping factor,
\begin{equation}
H(t) = -i~{\rm sgn(t)}~F(t)~e^{-{\delta}|t|}
\label{eq:KK4}
\end{equation}

The disadvantage of using this method is the same as that which arises
when trying to evaluate the KK integral directly, namely, the need to
have function values for frequencies more than twice the range of
frequencies of interest.  This is particularly troublesome for
functions that don't quickly die down within the energy range of
interest.  In our problem this requires the values of the momentum
matrix elements for states up to 30 eV above the Fermi level.  In the
following section we shall show our results of using this method for
the elements of the conductivity tensor in the optical as well as X-ray
region and compare it with non-KK transformed results.

We have applied the results of the foregoing analysis to determine
both the polar and the equatorial Kerr effects.  For the polar Kerr
effect the complex Kerr rotation is given by the relation,\cite{Reim}
\begin{equation}
\phi = \frac{-\sigma_{xy}}{\sigma_{xx}\sqrt{1 +
4{\pi}i\sigma_{xx}/\omega}}
\label{eq:polKerr}
\end{equation}

We have calculated the equatorial Kerr effect in the optical as well
as the soft X-ray region.  It is however well known that in the
equatorial geometry, the reflection coefficient of $p$-polarized light
at oblique incidence depends on the direction of magnetization because
of its dependence on the off-diagonal component of the conductivity
tensor.  Thus it is evident that a reversal in magnetization should
cause a change in the absorbed intensity of $p$-polarized light.
Moreover, since such an effect does not occur for $s$-polarized light
the equatorial Kerr effect can be observed\cite{Freiser} even using
unpolarized light.  Past calculations\cite{Freiser,Florczak} for the
equatorial Kerr effect in the optical region have used expressions that
are correct up to first order in $\kappa_2$, the off-diagonal component
of the dielectric tensor.  To calculate the effect in the X-ray region
we need the exact expression for the reflection coefficient of
$p$-polarized light incident at an angle of incidence $\theta$,
\begin{equation}
r = \frac{\cos\theta \biggl[n\kappa_{1}\beta +
\kappa_{2}\sin\theta\biggr] +
\sin^{2}\theta - \kappa_1}{ \sin^{2}\theta - \kappa_1 - \cos\theta
\biggl[n\kappa_{1}\beta + \kappa_{2}\sin\theta\biggr]}
\label{eq:eqKerr}
\end{equation}
where $\beta^{2} = 1 - \sin^{2}\theta/n^2$ and the complex refractive
index $n$ is $n^{2} = \kappa_{1} + \kappa_{2}^{2}/\kappa_{1}$ where
$\kappa_{1}$ is the diagonal element of the dielectric tensor.  From
Eq.~(\ref{eq:eqKerr}) the absorption $(1 - |r|^{2})$ can be obtained.

Lastly, we also calculated the X-ray Faraday rotation for nickel and
iron at the $2p$ edge.  Here we have used the standard expression for
the Faraday rotation $\theta_{F}$ for a thickness $d$,
\begin{equation}
\theta_{F} = {\frac{{\omega}d}{2c}}Re\biggl[n^{r} - n^{l}\biggr]
\label{eq:eqFar}
\end{equation}
where $n^{r,l} =\sqrt{1 + 4{\pi}i\sigma^{r,l}(\omega)/\omega}$ and
$\sigma^{r,l}(\omega) = \sigma_{xx}(\omega) \pm i\sigma_{xy}(\omega)$.

\section{Results and Discussion}

\subsection{Elements of the conductivity tensor}

Using the results of the band structure calculation, we calculated
$\sigma_{xx}^1$\/($\omega$) and $\sigma_{xy}^2$\/($\omega$) in the
optical as well as the X-ray region for nickel and iron.  For the
optical region, we show our results in comparison with experiments and
the theoretical results of Ref.~\onlinecite{Oppeneer1}. For comparison
purposes we have not included the Drude term to the diagonal terms in
the figures for $\sigma_{xx}$\/($\omega$).  Later, they have been
included in the calculation of the magneto-optical effects. Fig.~1(a)
shows the results  of  $\sigma_{xx}^1$\/($\omega$) for nickel.  Our
theoretical curve seems to be in good agreement with the experiments of
Ref.~\onlinecite{Ehren}.  The difference between our results and those
of Ref.~\onlinecite{Oppeneer1} is due to a different choice for the
lifetime.  After repeating the calculation with different values of the
lifetime parameter, we chose a value of 0.0368~Ry for nickel to give
best results for MOKE.  We also observe the 1~eV shift at 5.4~eV which
has been ascribed to the failure of the LDA in producing some nickel
$3d$ bands.\cite{Liebsch}  In the case of
$\omega\sigma_{xy}^2$\/($\omega$) (Fig.~1(b)) our results are similar
to those of Ref.~\onlinecite{Oppeneer1} but the peaks are more
pronounced in our case.  Another feature worth noting is the dip near
5.5~eV that is closer to the observed dip in our calculation than
theirs.  This has a very noticeable effect on the Kerr angle spectra as
we shall see later.  In the case of iron, our theoretical results for
$\sigma_{xx}^1$\/($\omega$) (Fig.~1(c)) compare far better with
experiment than those of nickel and agrees well with the results of
Ref.~\onlinecite{Oppeneer1}.  Our theoretical results seem to be closer
to the experiments of Ref.~\onlinecite{wea} and agreement in general
can be taken to be quite good.  In the case of
$\omega\sigma_{xy}^2$\/($\omega$) (Fig.~1(d)) our theoretical curve
displays a peak at 2.7~eV, that is noticeably higher than the
experimentally observed peak.  Here, however, the overestimation is
surprising since we used an inverse life time of 0.06~Ry for iron,
higher than the 0.05~Ry used by Ref.~\onlinecite{Oppeneer1}.

Employing the KK transform method outlined earlier, we calculated
$\sigma_{xx}^2$\/($\omega$) and $\sigma_{xy}^1$\/($\omega$).
Figure~2(a) shows the result for $\omega\sigma_{xx}^2$\/($\omega$) in
the case of nickel while Fig.~2(b) shows the one for iron.  In both
cases, our results are remarkably close to the results of
Ref.~\onlinecite{Oppeneer1}. In the case of nickel they also compare
well with experiments.\cite{johnson}  In the case of iron the two
experimental results seem to differ quite widely above 2~eV, thus
making comparison with theory quite difficult.  Figures~2(c) and
Fig.~2(d) show our results for $\omega\sigma_{xy}^1$\/($\omega$).  For
nickel, the disagreement at 5~eV is very obvious.  In our calculations,
this does not seem to have affected the polar Kerr angles as seriously
as it has the equatorial Kerr effect in the 0-10~eV region.  Other than
this, the results for nickel seem to agree very well with the general
features of the experiment.  For iron, the results of
$\omega\sigma_{xy}^1$\/($\omega$) follow the experiment quite well and
also compare well with the results of Ref.~\onlinecite{Oppeneer1}.

\subsection{MOKE and X-MCD}

Using the calculated curves for the elements of the conductivity tensor
we evaluated the Kerr angles for nickel and iron for the polar Kerr
geometry using Eq.~(\ref{eq:polKerr}).  The resulting curves for the
optical region are shown in Fig.~3(a) and Fig.~3(b).  In the
calculation of the Kerr angles, we have included the effect of a
phenomenological Drude term using values for $\sigma_{D}$ and
$\tau_{D}$ from previous experimental results.\cite{Lenham} The results
on nickel are particularly good since the shift of 1~eV is not as
noticeable in the Kerr angle spectra as it is in the elements of the
conductivity tensor.  This is in slight contrast to the observations of
Ref.~\onlinecite{Oppeneer1}.

In a later publication\cite{Oppeneer2} Oppeneer {\em et al} have
investigated the dependence of MOKE spectra on the strength of the
spin-orbit coupling.  They concluded that the MOKE peaks scale linearly
with the spin orbit coupling parameter $\xi$ and varying $\xi$ could
produce a better agreement with the observed MOKE spectra for nickel.
No such adjustments for $\xi$ were necessary in our results and our
theoretical results for nickel seem to agree very well with the
experimental results of Ref.~\onlinecite{Krinchik} throughout the
energy range of the data.  In the light of these observations we may
conclude that the spin-orbit coupling strength in our $d$-bands is
quite accurate.  In the case of iron, our Kerr angles are in excellent
agreement with experimental results\cite{Krinchik} and also compare
well with the theoretical results of Ref.~\onlinecite{Oppeneer1} and
Ref.~\onlinecite{Guo1}.

As noted before, one of the features of our KK transformation method
is that for functions that do not decay to zero within the required
energy range of interest, accurate momentum matrix elements for
energies up to two times as much are needed to correctly produce the KK
transformation.  Our momentum matrix elements are sufficiently accurate
up to those energies since they are calculated using simple analytic
expressions resulting from the use of a Gaussian basis and hence are
free of any numerical approximations.  In addition, it is evident from
our results of Kerr angles that with a straightforward inclusion of
spin-orbit coupling (for which we again have analytic expressions) in a
manner described in the earlier section, we are able to very
effectively and efficiently account for all the principal features of
the MOKE spectra.

It is interesting to see whether the strength of the spin-orbit
coupling in our $d$-bands that has given us good results for MOKE,
gives consistent results for the other magneto-optical properties as
well.  This is a very important test since it is now widely
accepted\cite{Oppeneer1,Guo1} that MOKE depends sensitively on the
strength of the spin-orbit coupling and the exchange splitting.  To
further investigate whether or not our spin-orbit coupling in the
valence bands is accurate we decided to calculate the X-ray MCD spectra
of iron and nickel at the $2p$ edge using our first principles LCGO
method.  This serves as a simultaneous check for the accuracy of the
LDA at X-ray energies.  In the X-MCD at the $2p$ core edge, it is known
that while the separation between the two peaks originates principally
from the spin-orbit splitting of the $2p$ levels, the exact ratio
between the two peaks (e.g. approximately -1.6:1 in the case of nickel)
arises out of the spin-orbit splitting of the $3d$ valence bands.
Since X-MCD is the difference in absorption of right and left
circularly polarized X-rays in the polar geometry, this is nothing but
$\sigma_{xy}^2$($\omega$) evaluated using momentum matrix elements
between core $2p$ and valence $3d$ bands.

The results of this calculation are shown in Fig.~4(a) for nickel and
Fig.~4(b) for iron.  Along with this, we also computed the
$\sigma_{xx}^1$\/($\omega$) which is a measure of the total absorption
of right and left circularly polarized X-rays.  These results are shown
in Fig.~5(a) and Fig.~5(b).  As in any core to band transition the
effect of the core hole has to be accounted for, which we did by
recalculating bands and eigenfunctions with an increased effective $Z$
value and then using the new energies and wavefunctions for the final
state to calculate the momentum matrix elements.  The resulting L$_3$
and L$_2$ peaks occur at 856~eV and 874~eV for nickel, which agree
fairly well with the observed energies of 853~eV and 871~eV
respectively.  Similarly, the calculated L$_3$ and L$_2$ peaks for iron
occur at 710~eV and 723~eV, which agree fairly well with the observed
peaks of 707~eV and 720~eV respectively.  However, our main results in
the X-MCD spectra are the L$_3$-to-L$_2$ ratios which are -1.56:1 in
the case of nickel and -1.2:1 in the case of iron. For nickel this
seems to agree very well with the observed ratio of -1.6:1.  For iron
the theoretical ratio is slightly smaller than the observed ratio.
However it agrees with the ratio for iron determined by other first
principle methods based on LDA.\cite{Guo3,Wu2} Thus, the discrepancy in
the L$_3$-to-L$_2$ ratio for iron is not an artifact of our method but
may be due to the failure of the one-electron band picture.  One aspect
that must be pointed out in the X-MCD spectra of nickel is the missing
peaks, often referred to as B and B$^\prime$, 4~eV away from the
principal peaks on the high energy side.  These have been ascribed to
many body effects in nickel by past authors\cite{Smith} and have been
reproduced by Jo and Sawatzky in a many-body calculation on
nickel.\cite{Jo}  They have not been reproducible by any one-electron
band calculation.  The results of our band-structure based calculation
of MOKE and X-MCD are then an indication that, within the LDA
framework, our method does give an accurate and consistent description
of these phenomena.

\subsection{The Equatorial Kerr effect and Photoabsorption X-MLD}

The other principal tool for magneto-optical studies on ferromagnetic
materials has been the equatorial Kerr effect.\cite {Krigorb,Brubaker}
As mentioned above, the reflection coefficient of an incident
electromagnetic wave in this geometry depends on the sense of
magnetization in the metal.  When observed at a core-level edge this
phenomenon may be termed X-ray magnetic linear dichroism.  It may be
noted that a different kind of absorption X-MLD can be obtained by
keeping the magnetization constant but rotating the photon polarization
vector by 90\/$^\circ$.  This effect has also been calculated
previously\cite{Guo3} but is not considered here.

We have calculated the equatorial Kerr effect for absorption both in
the optical region and in the soft X-ray region (X-MLD) using
Eq.~(\ref{eq:eqKerr}) for several different angles of incidence.  The
results for the optical region are shown Fig.~6(a) and Fig.~6(b) for
nickel and iron respectively.  From Fig.~6(a) it can be seen that as
the angle of incidence is changed from 45\/$^\circ$ to 80\/$^\circ$ on
nickel the sign of the effect is reversed.  After this, from
80\/$^\circ$ to 88\/$^\circ$ the peak magnitudes progressively increase
reaching a peak somewhere close to grazing incidence.  The effect, of
course, disappears at exactly 90\/$^\circ$.  For nickel, it is
important to remember that in the evaluation of the conductivity
tensor, the theoretical results have always predicted a dip at 5~eV
where there actually is an experimental rise (see Fig.~1(b) and
Fig.~1(d)).  Taking a clue from this observation, we may predict that
in Fig.~6(a), although the structure up to 4~eV may compare well with
experiment, the peaks (or dips) at about 5~eV may well be found to be
reversed for every angle.

The curves of Fig.~6(b) for iron can however be considered to be
faithfully reproducing experimental results.  For iron, we see that as
the angle is increased from 80\/$^\circ$ to 85\/$^\circ$, there is a
reversal of the sign of the effect only for the region upto 5~eV.  The
effect is most enhanced at 88\/$^\circ$ to 89\/$^\circ$ and the most
observable change is at about 6~eV.  Although we did not find any
experimental results for change in absorption upon reversal of
magnetization in the optical regime, experimental results for the
change in the reflection intensity between magnetized and unmagnetized
nickel have been reported in the past.\cite{Krigorb}  We calculated
this also using Eq.~(\ref{eq:eqKerr}) by putting $\sigma_{xy}$ equal to
zero to simulate the unmagnetized nickel.  Although this procedure may
be viewed with caution, our results for nickel as shown in Fig.~6(c)
agree remarkably well with experiment.  There again is the
characteristic dip at 5~eV which does not follow the rise seen in
experiment, but this has already been noted to be a consistent failure
of LDA in nickel.

For the X-ray region our results for the $2p$ edge have been shown for
nickel and iron in Fig.~7(a) and Fig.~7(b) respectively.  We clearly
see that at the onset of the L$_{2,3}$ edges in both cases the sign of
the peaks is the same.  There is, however, a  sharp overshoot to the
opposite sign just after the L$_2$ edge.  This is more pronounced in
the case of nickel, although it is unmistakable even in iron.  It would
be interesting to see if these features are actually observed.  In the
case of iron, the predicted peaks at the L$_3$ and L$_2$ edge seem to
be of almost equal magnitude.  This may be in slight disagreement with
experimental results since, as mentioned earlier, the prediction of the
L$_3$-to-L$_2$ ratios both in the X-MCD and total X-ray absorption
calculation are slightly underestimated for iron.  It may also be noted
that at the incident angles considered the maximum effect for iron is
about 0.8\% while that for nickel is about 1\%.  This agrees with the
analysis about this type of linear dichroism done in the past.
\cite{Guo3} To the best of our knowledge, such absorption X-MLD
experiments at the $2p$ edge for iron and nickel have not been reported
although X-MLD in photoemission has been subjected to an onslaught of
theoretical and experimental analysis.
\cite{Laan,Henk,Sirotti,Kuch,Tamura,Roth}  The results there are
characteristically different from our absorption results, since the
final states in a photoemission experiment are very high above the $3d$
bands.  As such, the final state spin-polarization does not play as
significant a role as it does in photoabsorption type measurements.
This is particularly demonstrated in photoemission-XMLD\cite{Roth}
where the signs of the two peaks are opposite each other, in contrast
to what is seen in our photoabsorption-X-MLD.

\subsection{Soft X-ray Faraday rotation}

We finally turn to the Faraday rotation of the plane of polarization
of linearly polarized X-rays upon transmission through magnetic
metals.  The main concern here stems from the knowledge that the
Eq.~(\ref{eq:eqFar}) derived for the Faraday effect is based on the
dipole approximation.  At a first glance, to use it for evaluating the
Faraday effect in the soft-X-ray regime is to bring the approximation
to a questionable limit.  This problem has been addressed
previously\cite{Gotsis} and it has been demonstrated that the theory of
Faraday effects can be extended to the X-ray regime by postulating the
existence of a free carrier effective dielectric medium with the same
conductivity tensor as the ferromagnetic metal in question.  In fact we
already did make this extension when we used $\sigma_{xy}^2$ to
determine X-MCD and $\sigma_{xx}^1$ to determine the total absorption.
Since those results produced reasonably good agreement with
experiments, we wish to examine its applicabilty further by evaluating
the Faraday effect at soft X-ray energies.  This calculation is a
trivial extension to the calculation of X-MCD and total absorption
since all we need is the KK tranforms of these curves.  This, as
mentioned earlier, can be done very efficiently and accurately with our
method.

The calculated Faraday rotation at the $2p$ edges of nickel and iron
is exhibited in Fig.~8(a) and Fig.~8(b).  An experiment on soft X-ray
Faraday rotation has recently been done at the $2p$ edge of
iron\cite{Kortright} with which we compare our results.  As can be seen
from Fig.~7(b) the agreement with experiment is rather good at the
L$_2$ edge but is relatively poor at the L$_3$ edge.  This again is
related to the underestimated peak in the MCD and total absorption
spectra at the L$_3$ edge of iron.  Nevertheless, given that the dipole
approximation is not strictly valid at these energies, the agreement
with observed angles can be taken to be quite good.  For the case of
nickel, because our X-MCD and total absorption results are in much
better agreement with experiments, we are more confident of the
relative magnitudes of angles at the L$_3$ and L$_2$ edges.

\section{Conclusions }

A first principles, self consistent LCGO band calculation has been
performed and the principal magneto-optical properties have been
computed for bulk iron and nickel in this paper.  In spite of the
simplistic way in which spin-orbit coupling has been included, our
results have agreed very well with previous first principles
calculations based on LDA of MOKE and MCD and experimental
observations.  In particular our MOKE results for nickel have produced
very good agreement with experiments, despite the lack of good
description of some of the LDA bands in nickel.  Discrepancies
originating from the failure of the LDA have however crept up in the
equatorial Kerr effect in the optical region around 5~eV.  Our results
for photoabsorption X-MLD could not be compared with experiments but
judging from the agreement of the X-MCD and soft X-ray rotation results
with experiments, we conclude that our results for nickel may be
considered to be accurate, but for iron the actual L$_3$ peak magnitude
may be much higher as compared to the L$_2$ peak.  This conclusion is
also supported by our Faraday effect results on iron.  It is also
noteworthy that the equatorial Kerr effect is larger in the optical
region than in the soft-X-ray region.

We have also demonstrated that fast, efficient and accurate KK
transforms can in fact give sufficiently satisfactory results for
sensitive magneto-optical effects of bulk metals, which have been
previously obtained by computationally intensive methods.  This speed
and efficiency is particularly important if one is to use such
calculations for quickly obtainable results for technological
applications.  This analysis has also confirmed that expressions for
the Faraday effect for the optical region may be extended to the soft
X-ray region yielding satisfactory results.  The principal feature of
this work is however that a single calculation of all the components of
the conductivity tensor in both the optical as well as the X-ray region
has yielded consistent results for a host of different magneto-optical
effects.

\acknowledgments

One of us (N.M) would like to acknowledge K. Subramanian for useful
discussions in the theory of the equatorial Kerr effect.  This research
was supported by the National Science Foundation under Grant
Nos.~NSF-DMR-9120166 and NSF-DMR--9408634.

\begin{figure}

\caption{
Elements of the conductivity tensor for nickel and iron.  In (a) and
(b), solid line with an inverse lifetime of 0.0368 Ry is our result.
Dashed line is the result of the calculation of
Ref.~\protect\onlinecite{Oppeneer1} with an inverse lifetime of 0.04
Ry.  In (c) and (d) solid line is our result for an inverse lifetime of
0.06 Ry.  Dashed line is the result of the calculation of
Ref.~\protect\onlinecite{Oppeneer1} for an inverse lifetime of 0.05
Ry.  (a) $\sigma_{xx}^1$ for nickel.  Circles are the experimental
results of Ref.~\protect\onlinecite{Ehren}.  Squares are the
experimental results of Ref.~\protect\onlinecite{Shiga}.  (b)
$\omega\sigma_{xy}^2$ for nickel.  Circles are the experimental results
of Ref.~\protect\onlinecite{Krinch2}.  (c) $\sigma_{xx}^1$ for iron.
Circles are the experimental results of Ref.~\protect\onlinecite{wea}.
Squares are the experimental results of Ref.~\protect\onlinecite{yol}.
(d) $\omega\sigma_{xy}^2$ for iron.  Circles are the experimental
results of Ref.~\protect\onlinecite{Krinch2}.  }

\label{condnife}

\end{figure}

\begin{figure}

\caption{ Dispersive components of the conductivity tensor for nickel
and iron.  (a) $\omega\sigma_{xx}^2$ for nickel.  Circles are the
experimental results of Ref.~\protect\onlinecite{johnson}.  (b)
$\omega\sigma_{xx}^2$ for iron.  Circles are the experimental results
of Ref.~\protect\onlinecite{johnson}.  Squares are the experimental
results of Ref.~\protect\onlinecite{yol}.  (c) $\omega\sigma_{xy}^1$
for nickel.  Circles are the experimental results of
Ref.~\protect\onlinecite{Krinch2}.  (d) $\omega\sigma_{xy}^1$ for
iron.  Circles are the experimental results of
Ref.~\protect\onlinecite{Krinch2}.  }

\label{dispnife}

\end{figure}

\begin{figure}

\caption{
Polar Kerr rotation. In both graphs, solid line is our result.  Dashed
line is the result of the calculation of
Ref.~\protect\onlinecite{Oppeneer1}.  (a) For nickel.  Circles are the
experimental results of Ref.~\protect\onlinecite{Krinchik}.  Squares
are the experimental results of Ref.~\protect\onlinecite{vn}.  (b) For
iron.  Circles are the experimental results of
Ref.~\protect\onlinecite{Krinchik}.  }

\label{pkerrnife}

\end{figure}

\begin{figure}

\caption{
X-MCD at the $2p$ edges for (a) nickel and (b) iron.
A Gaussian broadening of 0.2~eV has been added in both
cases.
}

\label{xmcdnife}

\end{figure}

\begin{figure}

\caption{
Total absorption at the $2p$ edges for (a) nickel and
(b) iron.  A Gaussian broadening of 0.2~eV has been added
in both cases.
}

\label{tabnife}

\end{figure}

\begin{figure}

\caption{
Equatorial Kerr effect.
For (a) and (b) $\Delta T$ is the change in absorption intensity
upon reversal in magnetization.  $T$ is the average absorption
of the two directions.  Plots are correct upto a negative sign.
(a) In absorption for nickel and
(b) In absorption for iron.
$\theta$~=~45$^\circ$ Solid line.
$\theta$~=~80$^\circ$ Dotted line.
$\theta$~=~85$^\circ$ Dashed line.
$\theta$~=~88$^\circ$ Dash-dotted line.
(c) $\theta$~=~45$^\circ$.
$\Delta I$ is the change in reflection intensity
upon magnetization.  $I$ is the reflection intensity
from unmagnetized nickel.
Solid line is our theoretical result.
Circles are the experimental results of
Ref.~\protect\onlinecite{Krinchik}.
}

\label{eqkerrnife}

\end{figure}

\begin{figure}

\caption{
Photoabsorption X-MLD at the $2p$ edges.
$\Delta T$ is the change in absorption intensity
upon reversal in magnetization.  $T$ is the average absorption
of the two directions.  Plots are correct upto a negative sign.
(a) for nickel.
$\theta$~=~85$^\circ$ Solid line.
$\theta$ = 87$^\circ$ Dotted line.
$\theta$ = 88$^\circ$ Dashed line.
(b) for iron.
$\theta$~=~80$^\circ$ Solid line.
$\theta$~=~85$^\circ$ Dotted line.
$\theta$~=~87$^\circ$ Dashed line.
}

\label{xmldnife}

\end{figure}

\begin{figure}

\caption{
Soft X-ray Faraday rotation at the $2p$ edges.
In both figures solid line is our theoretical
results.
(a) for nickel.
(b) for iron, $d$~=~$80$~nm.
Circles are the experimental results of
Ref.~\protect\onlinecite{Kortright}.
}

\label{farnife}

\end{figure}

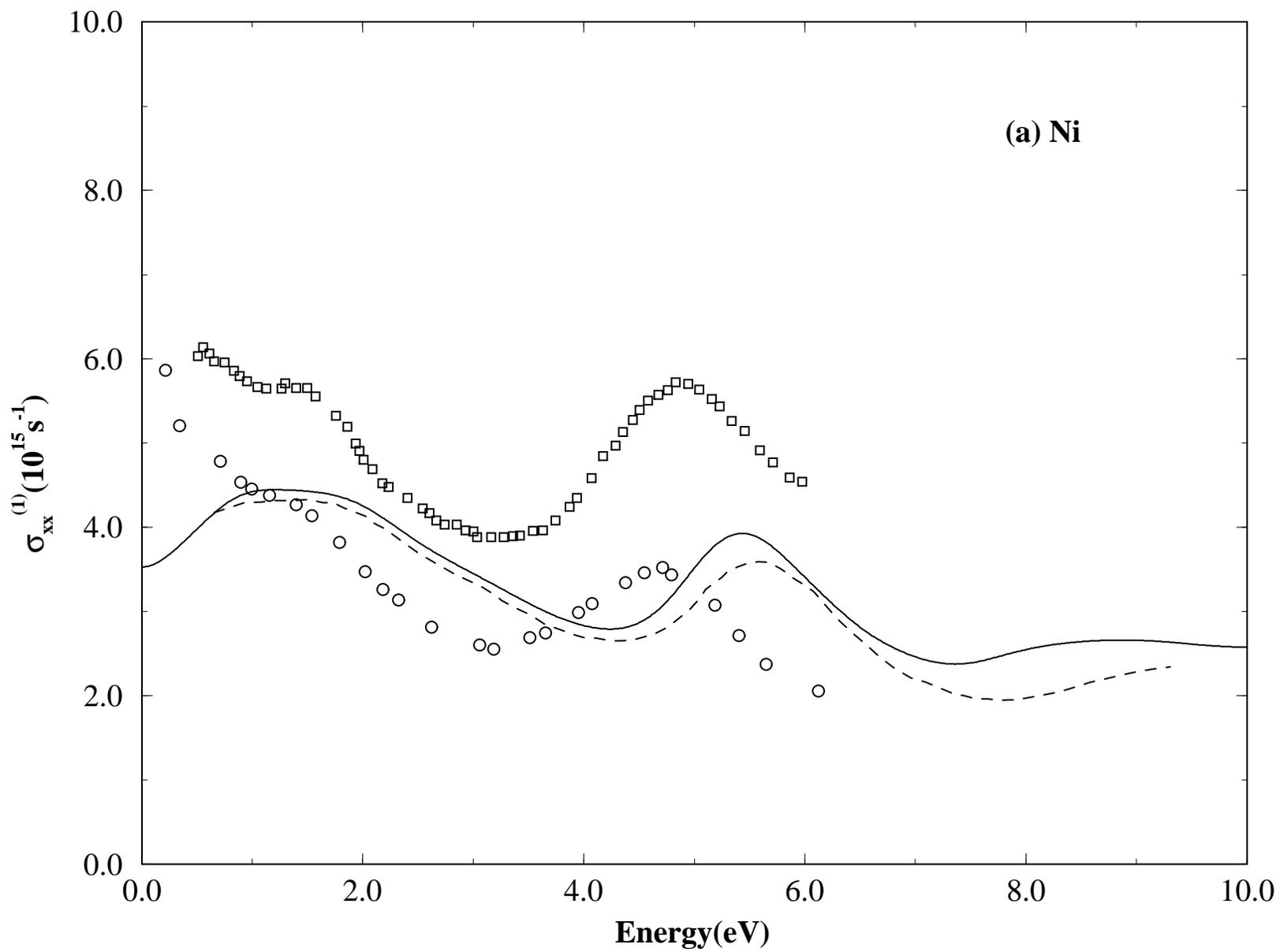
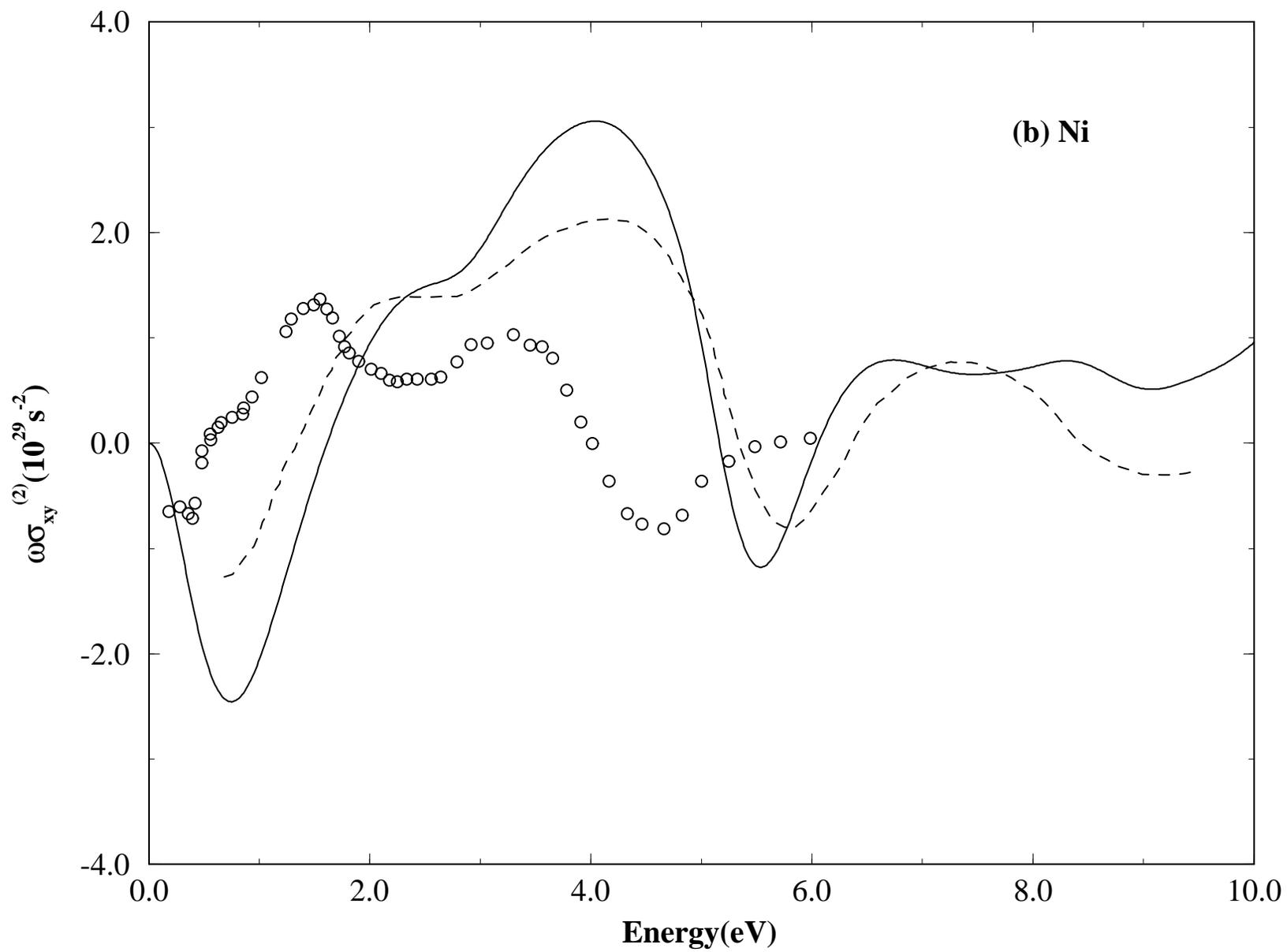
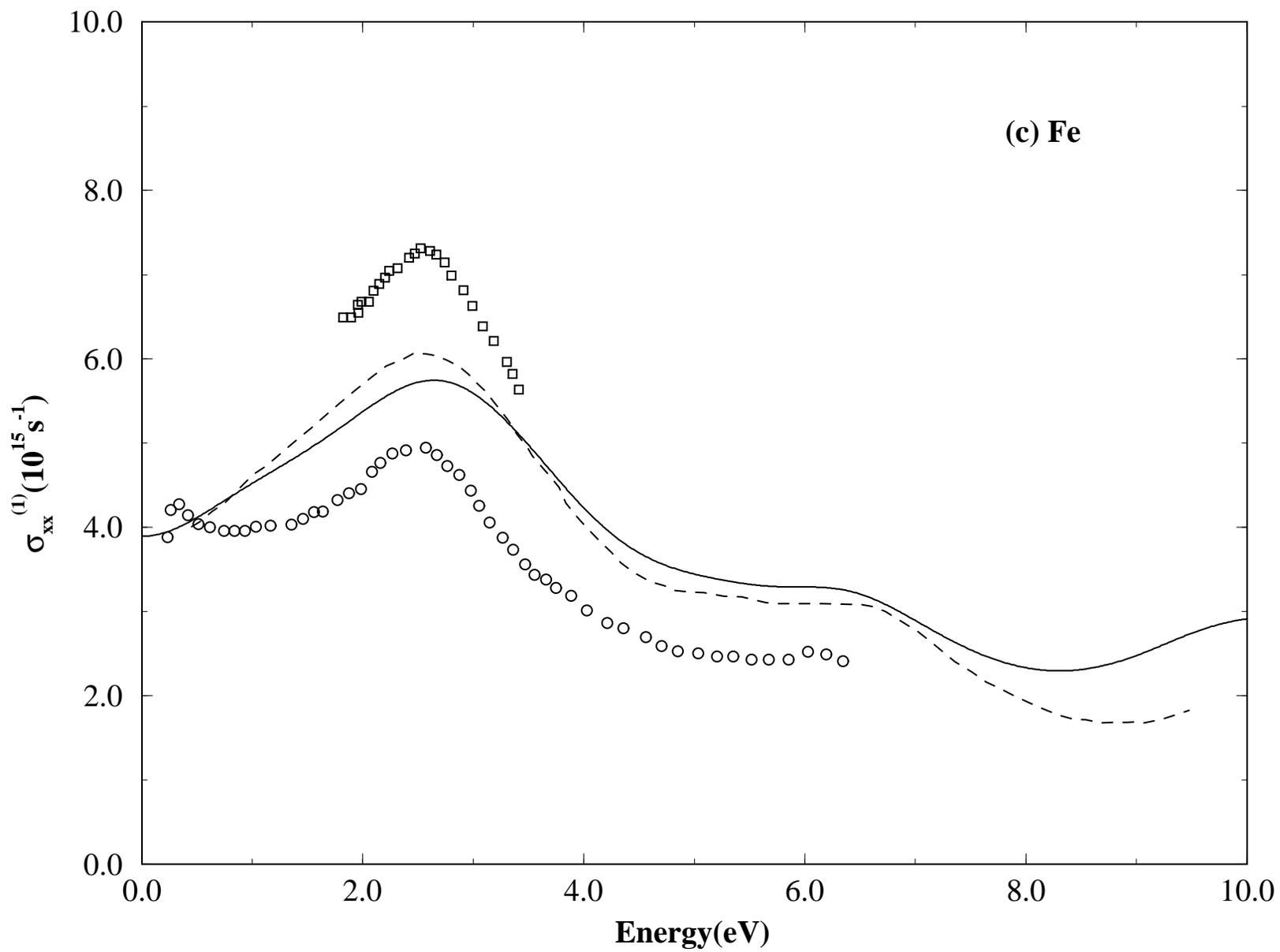
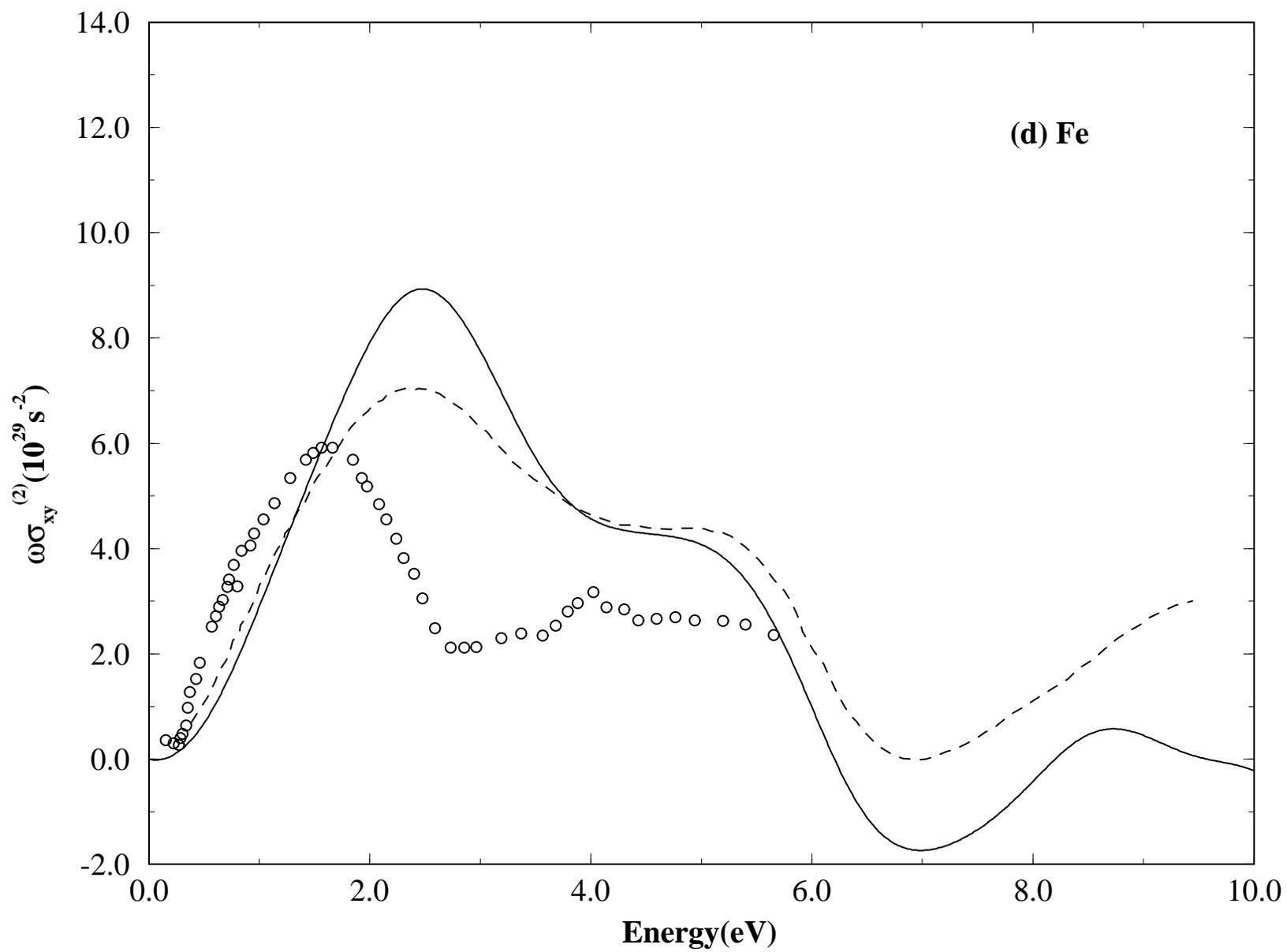
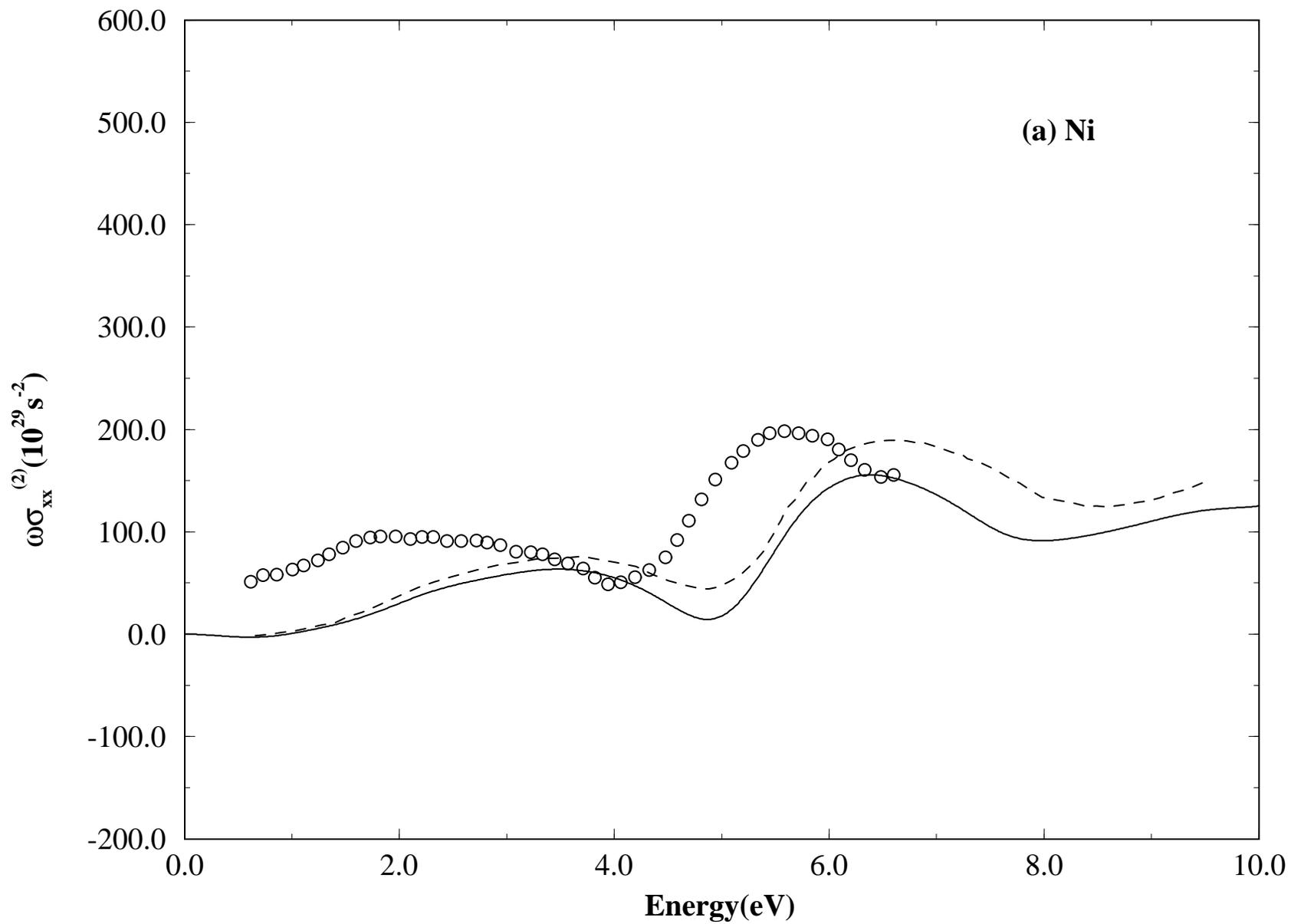
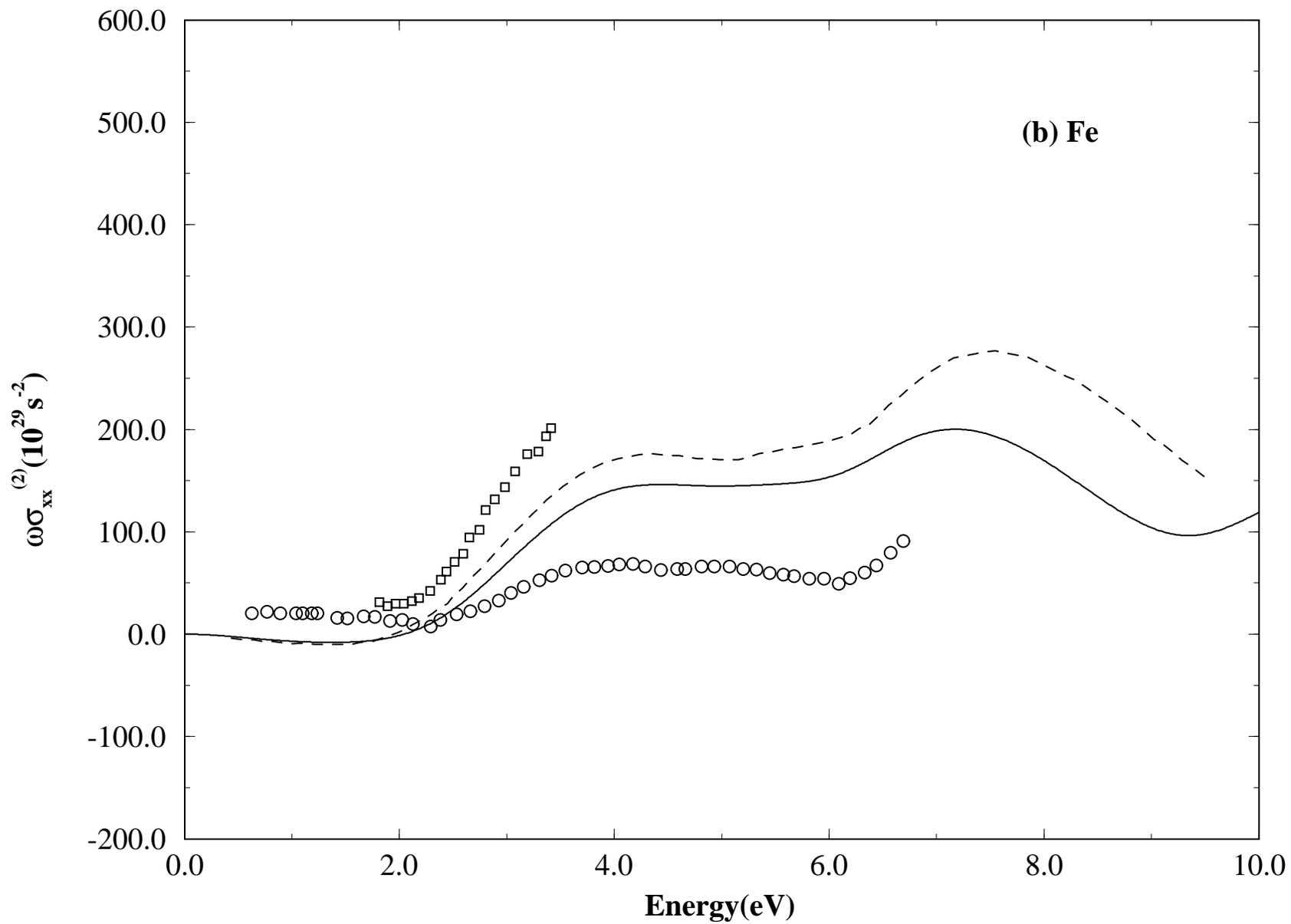
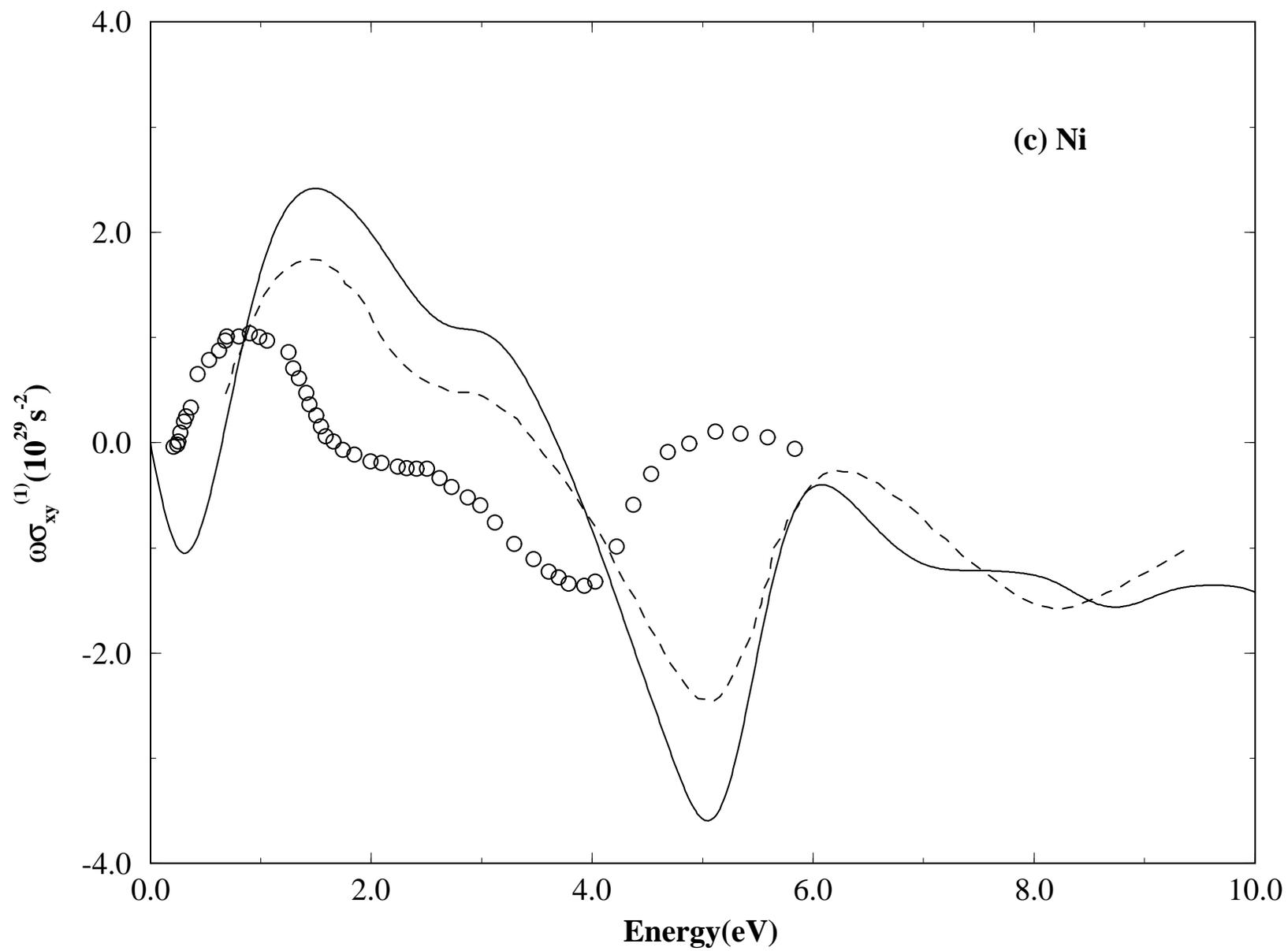
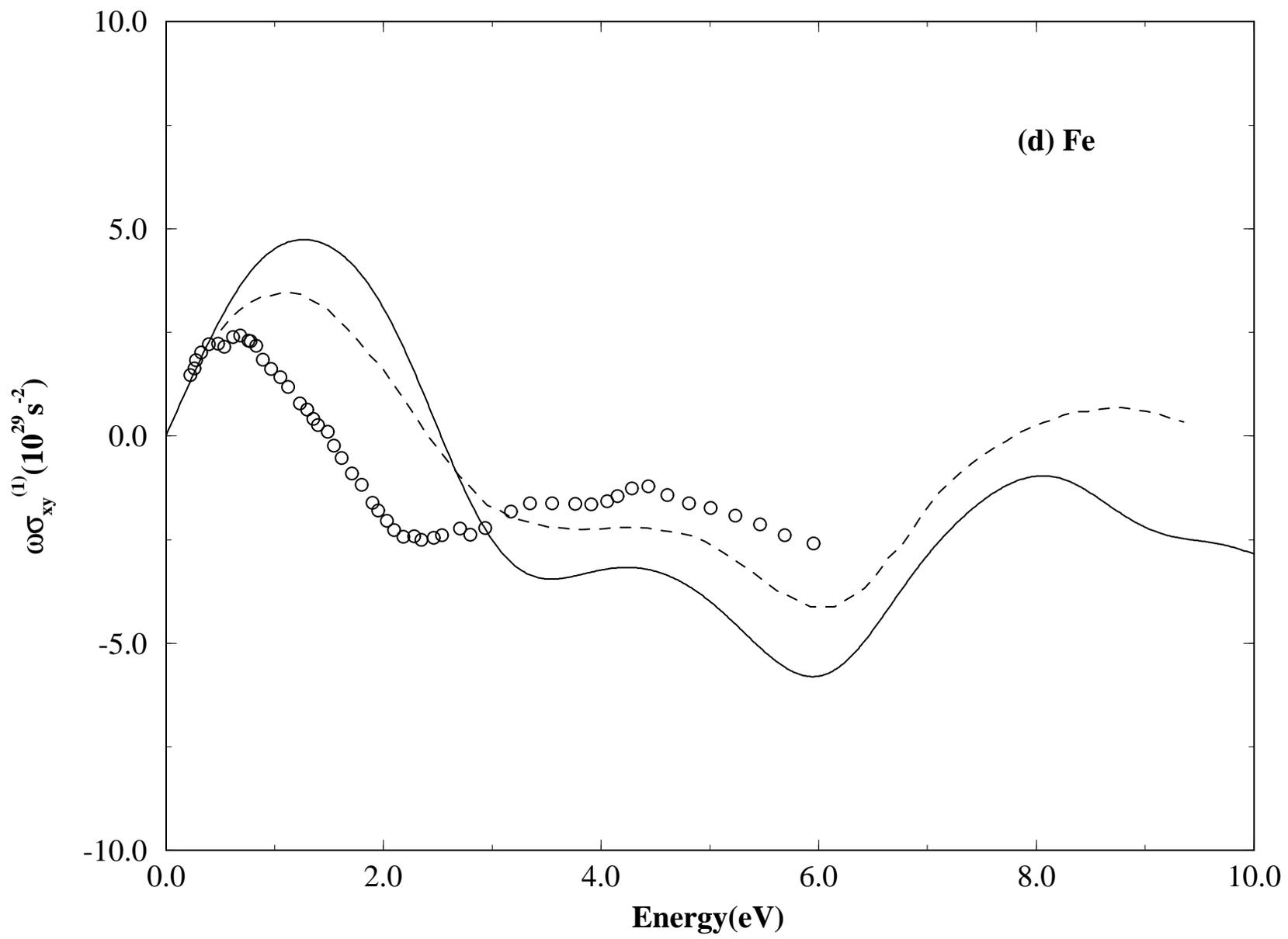
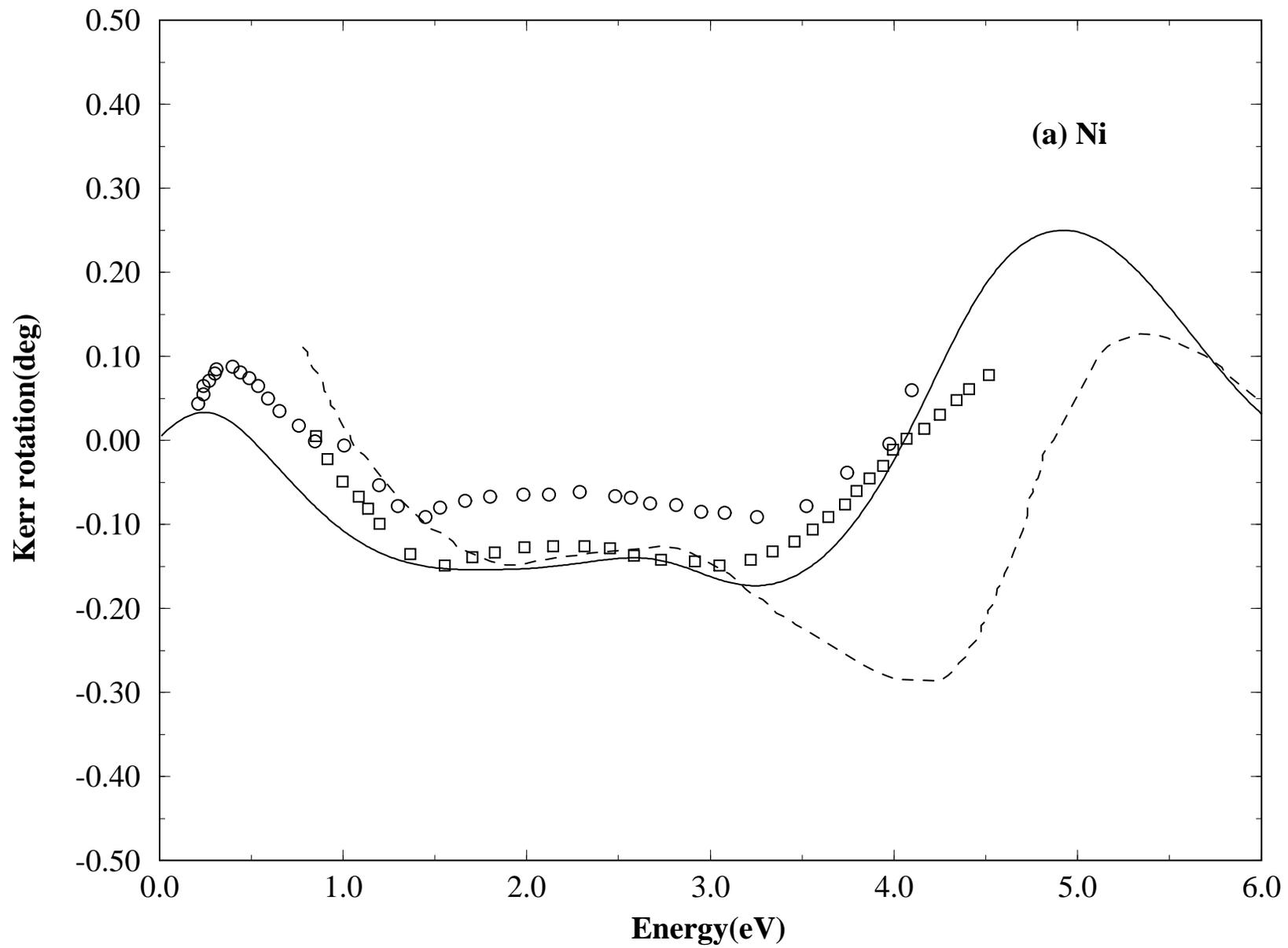
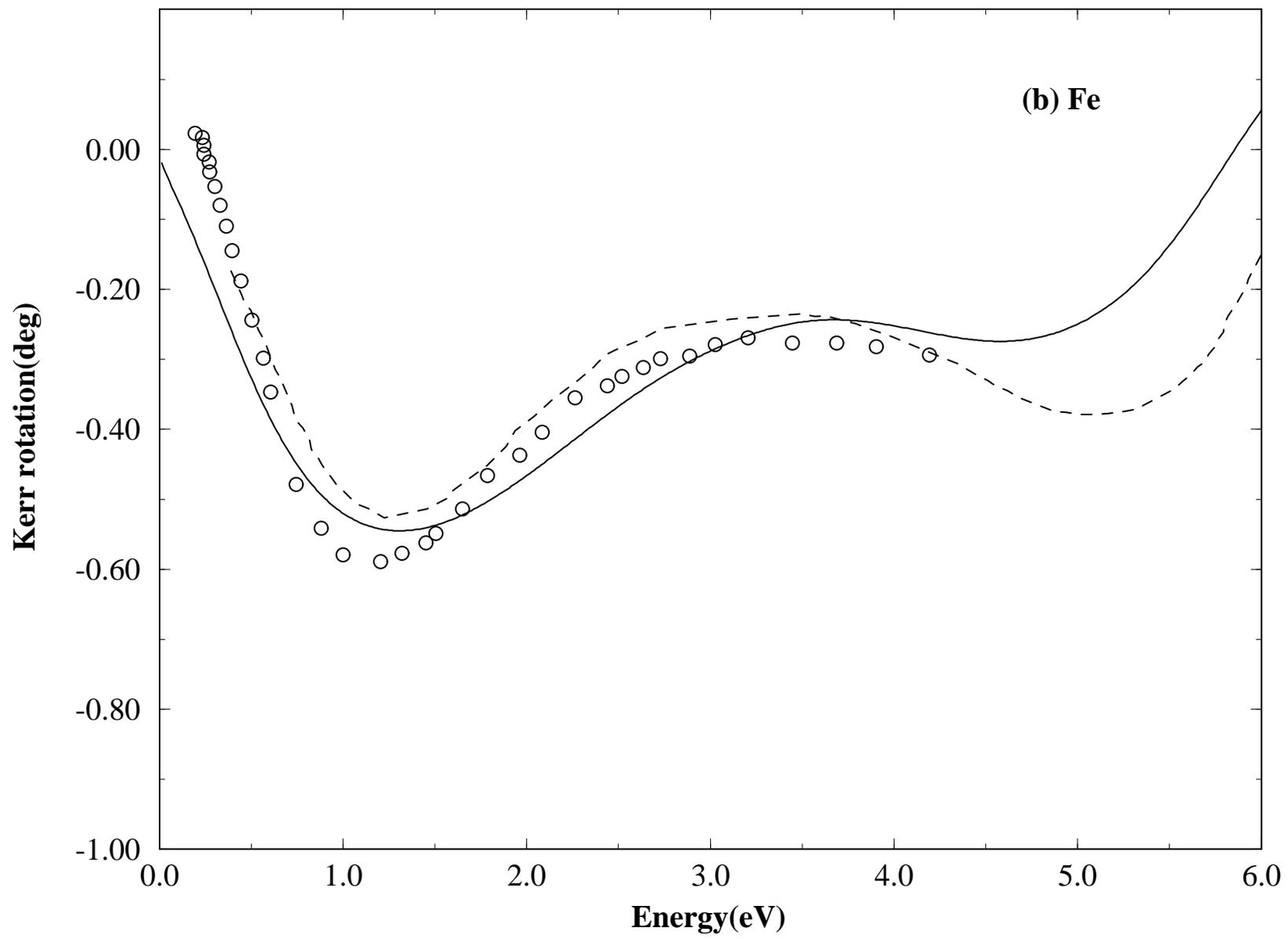
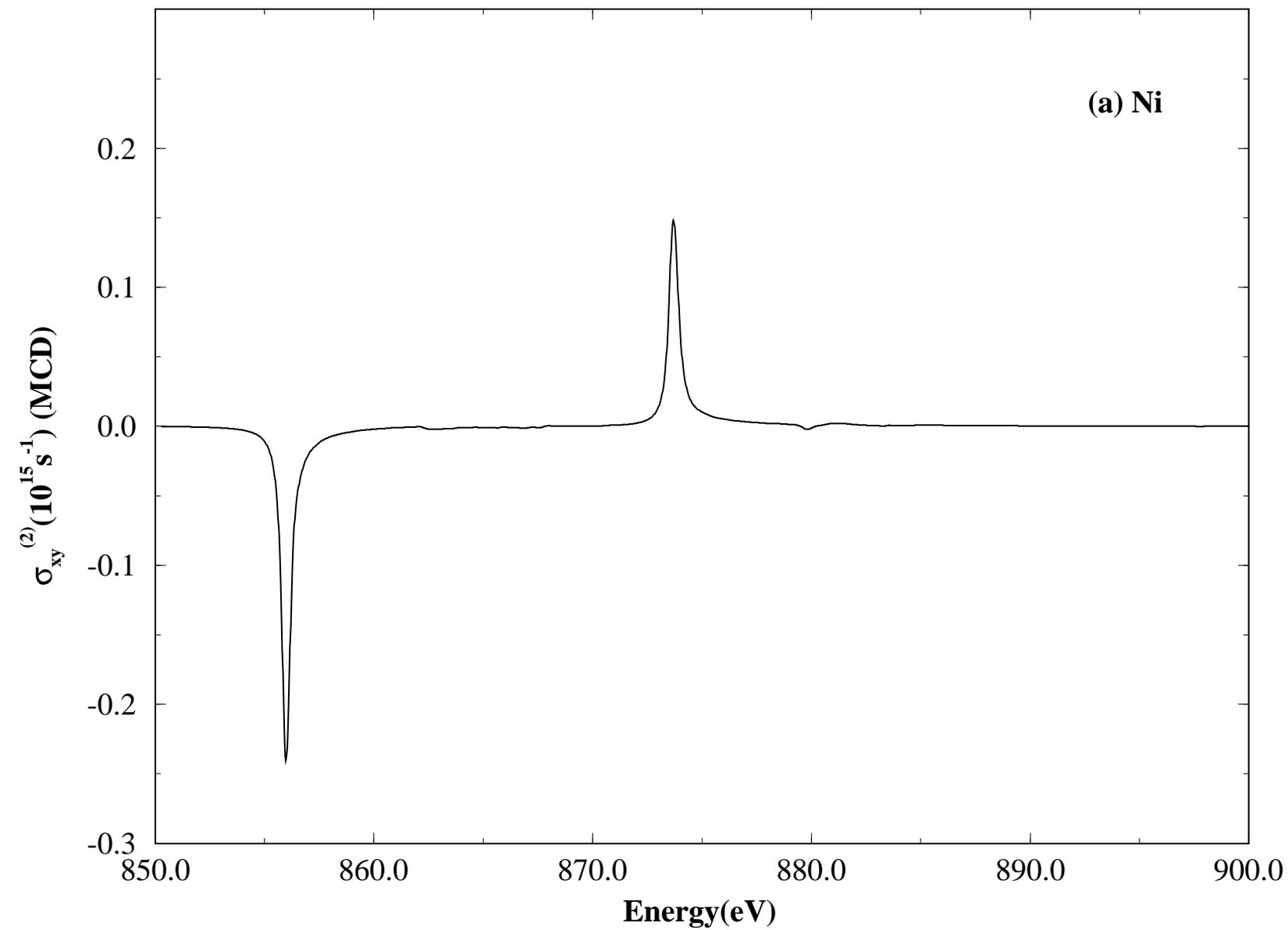
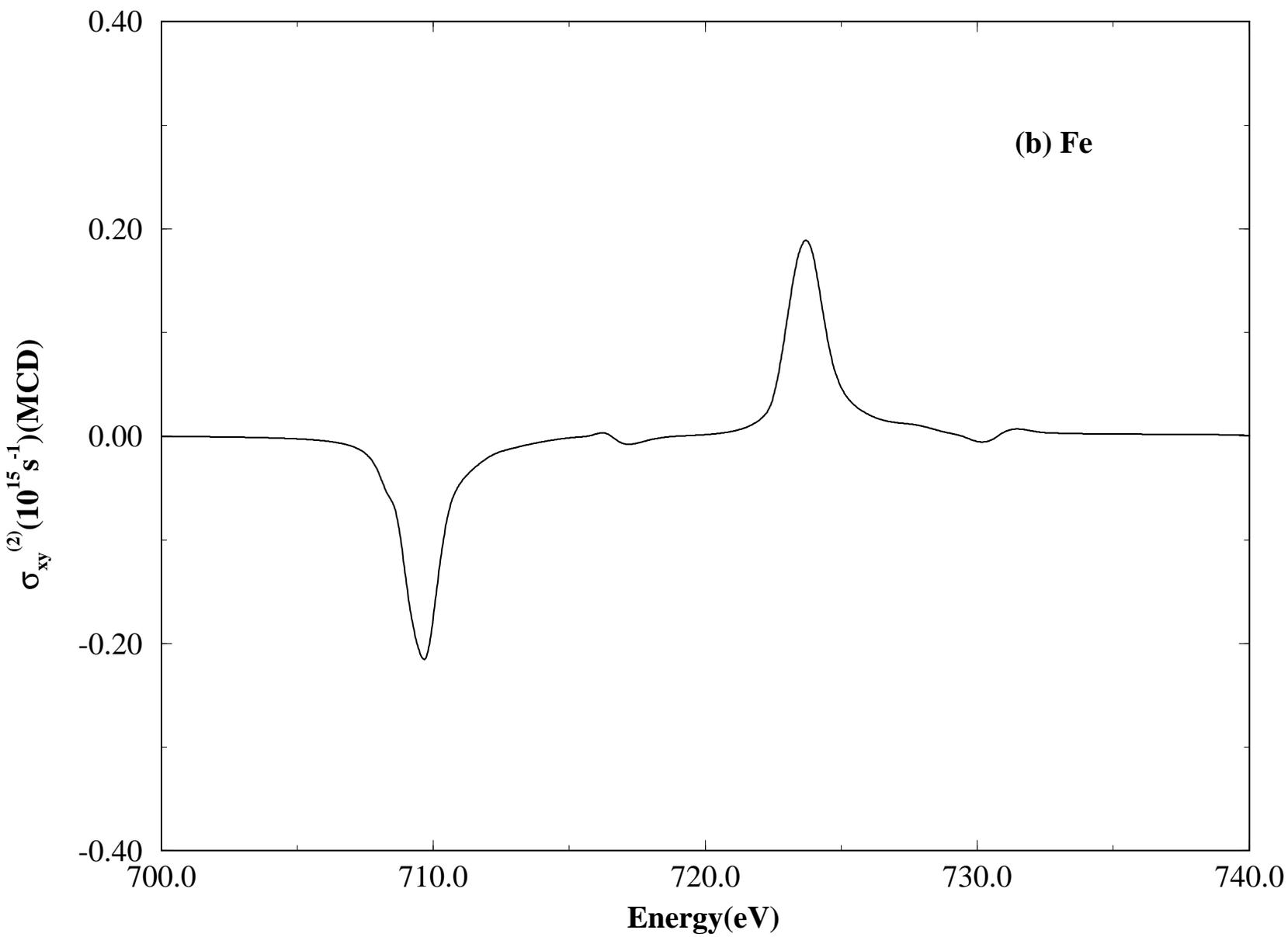
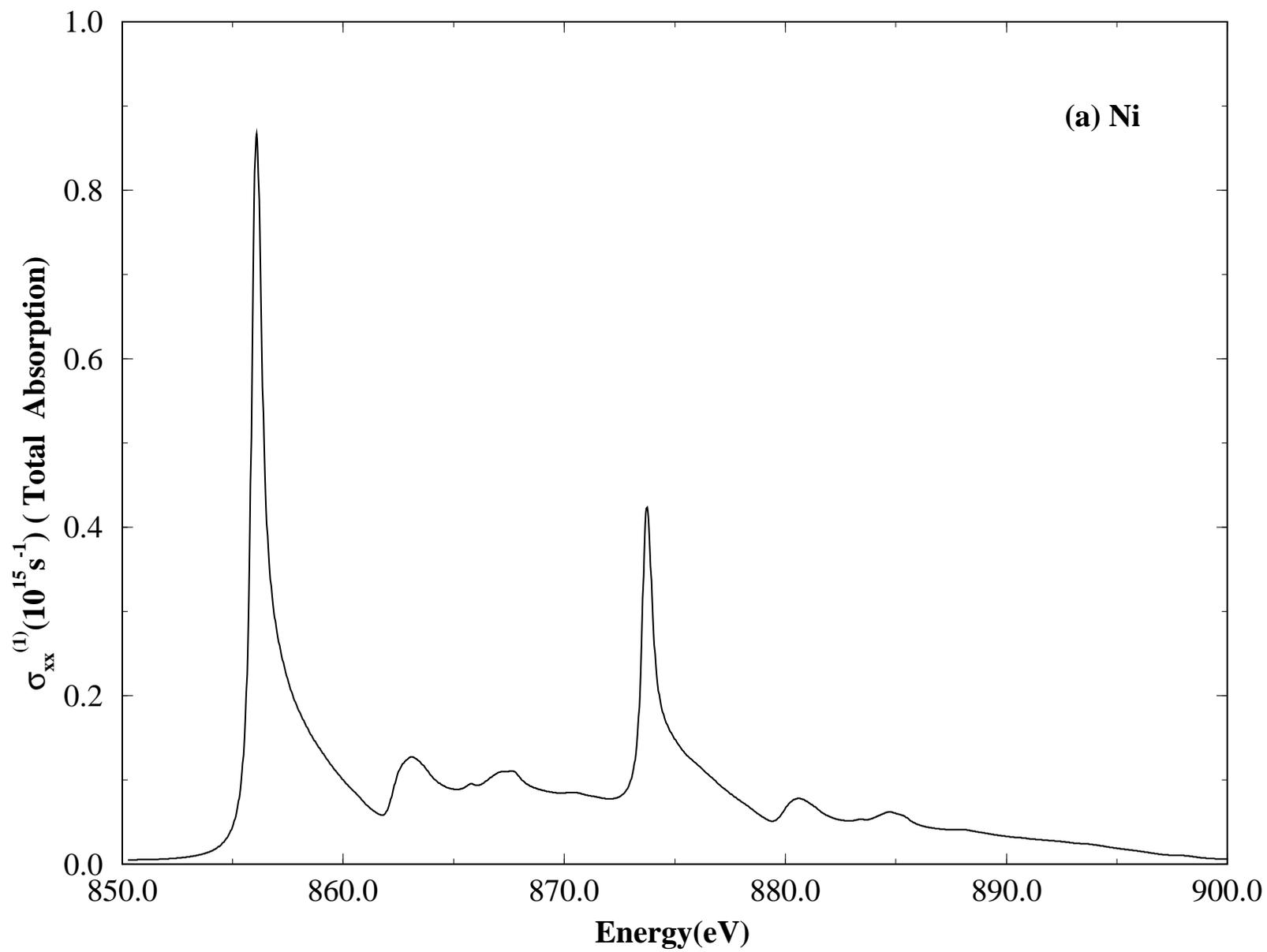
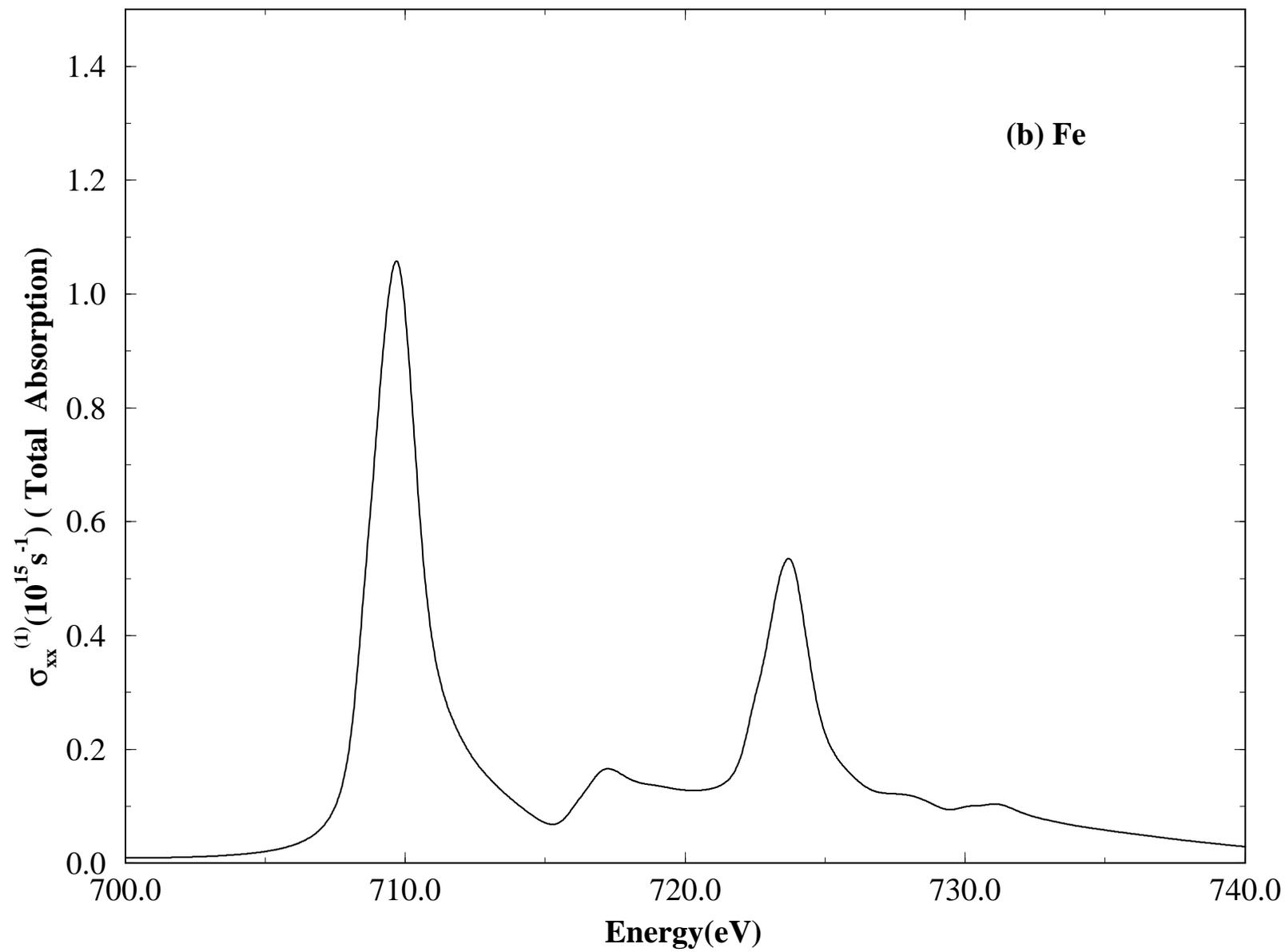
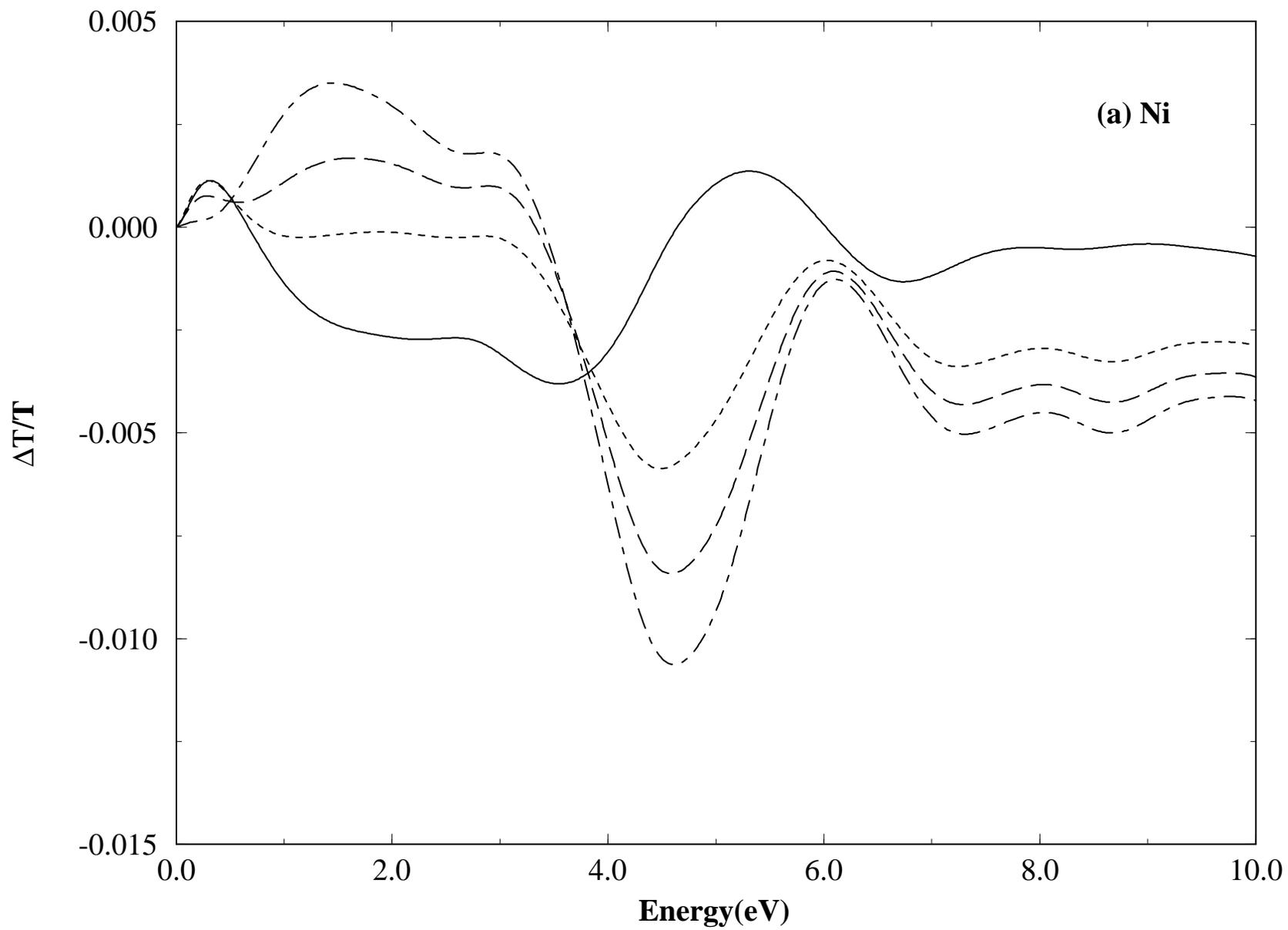
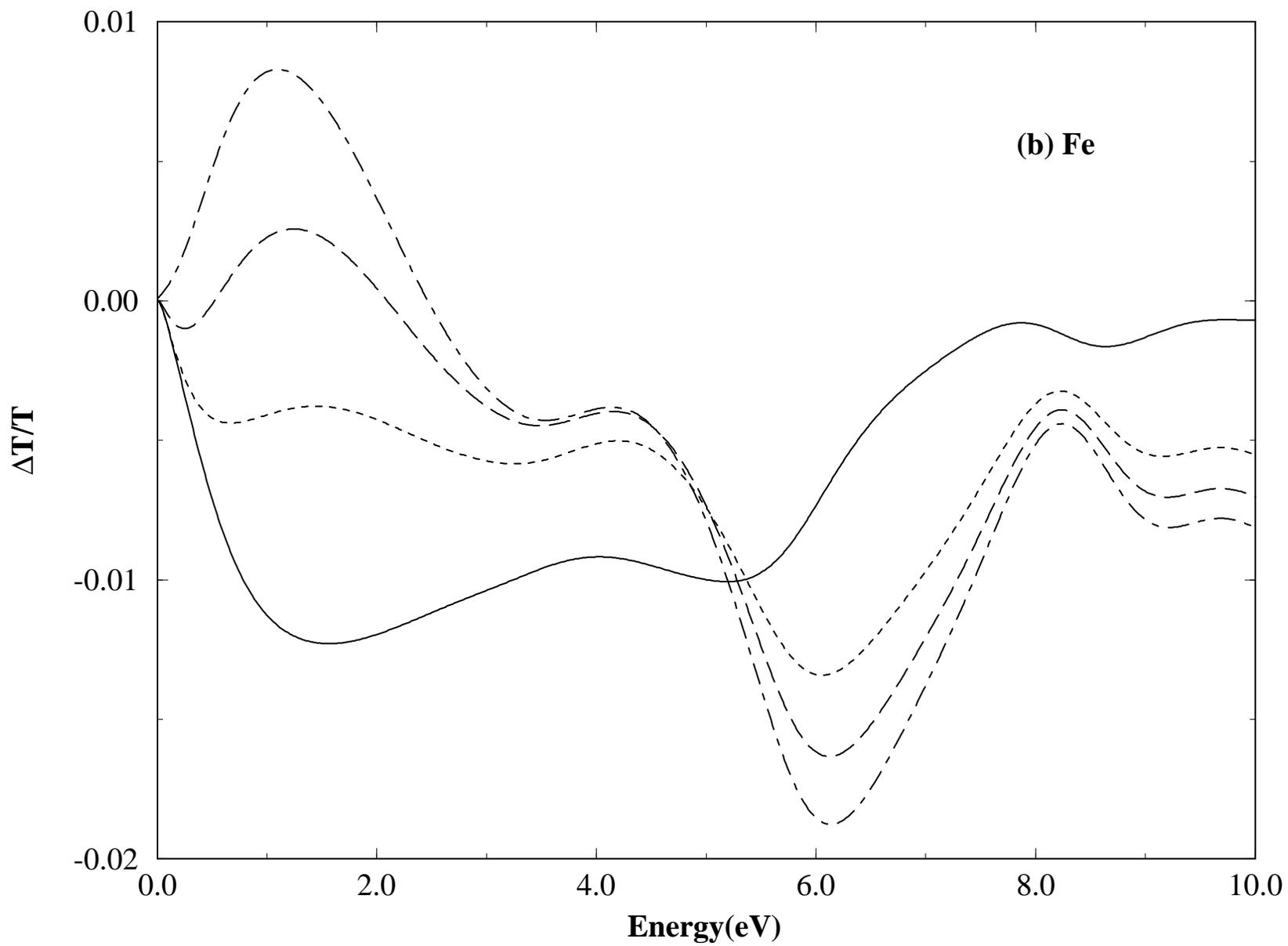
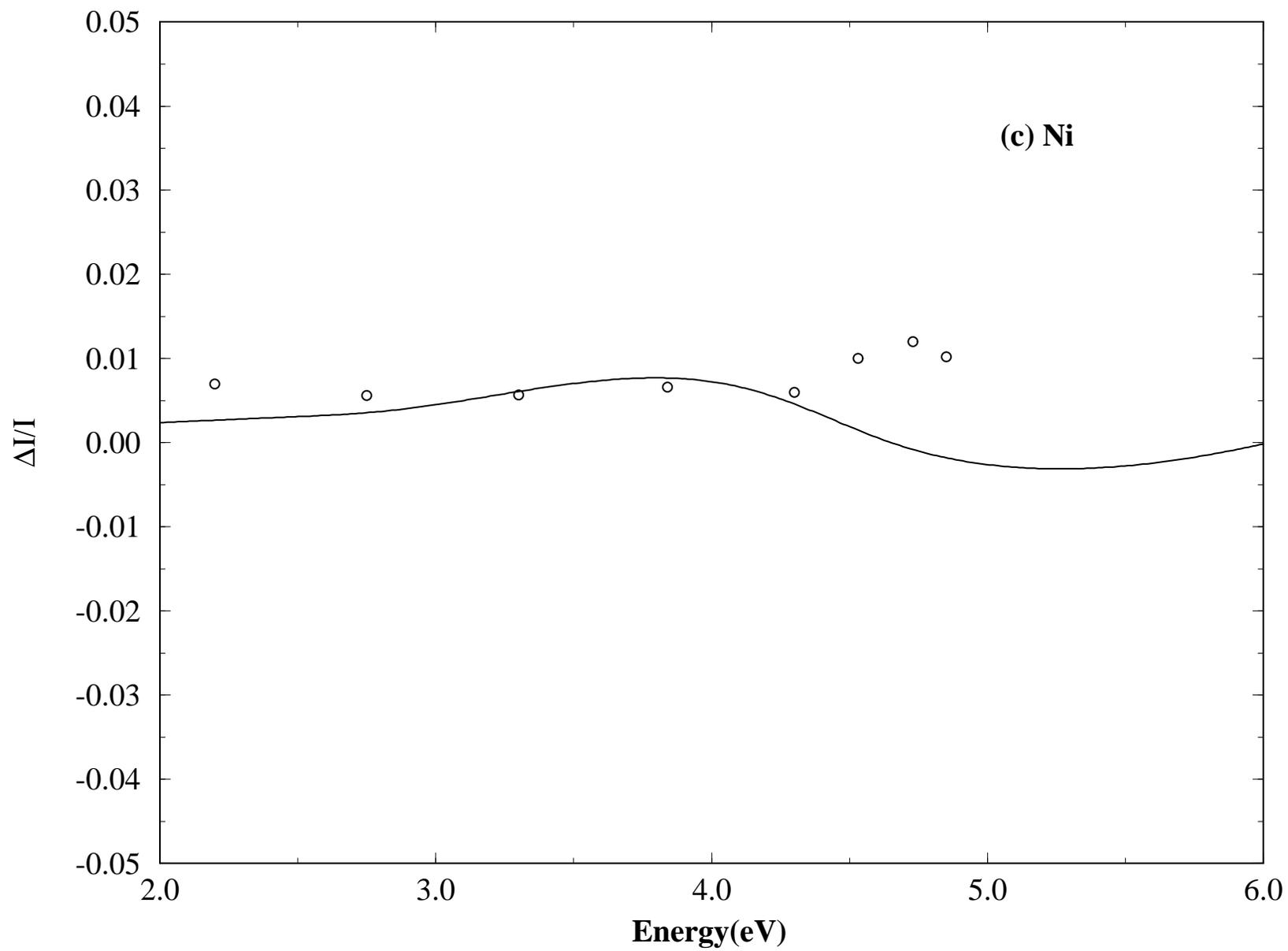
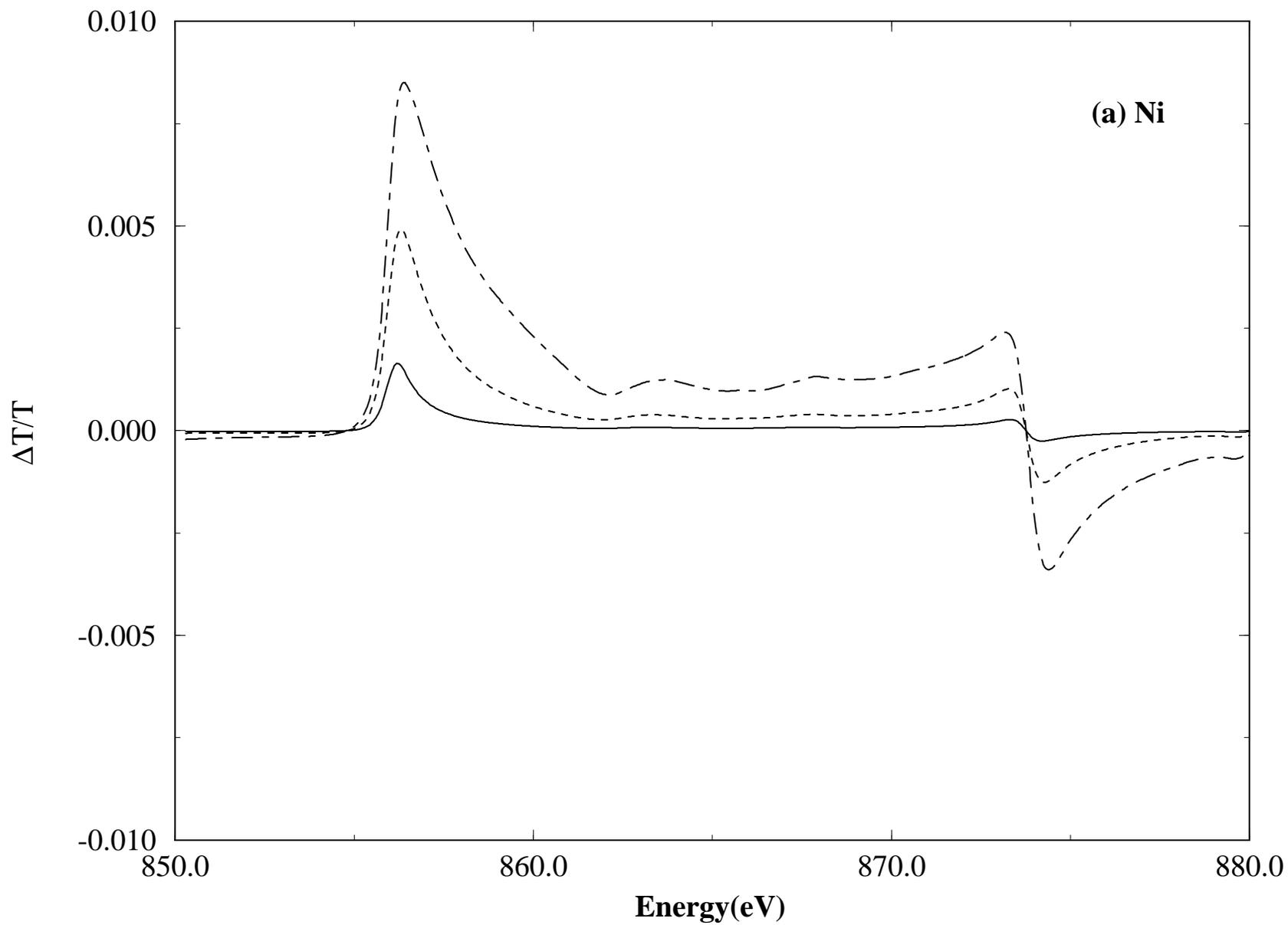
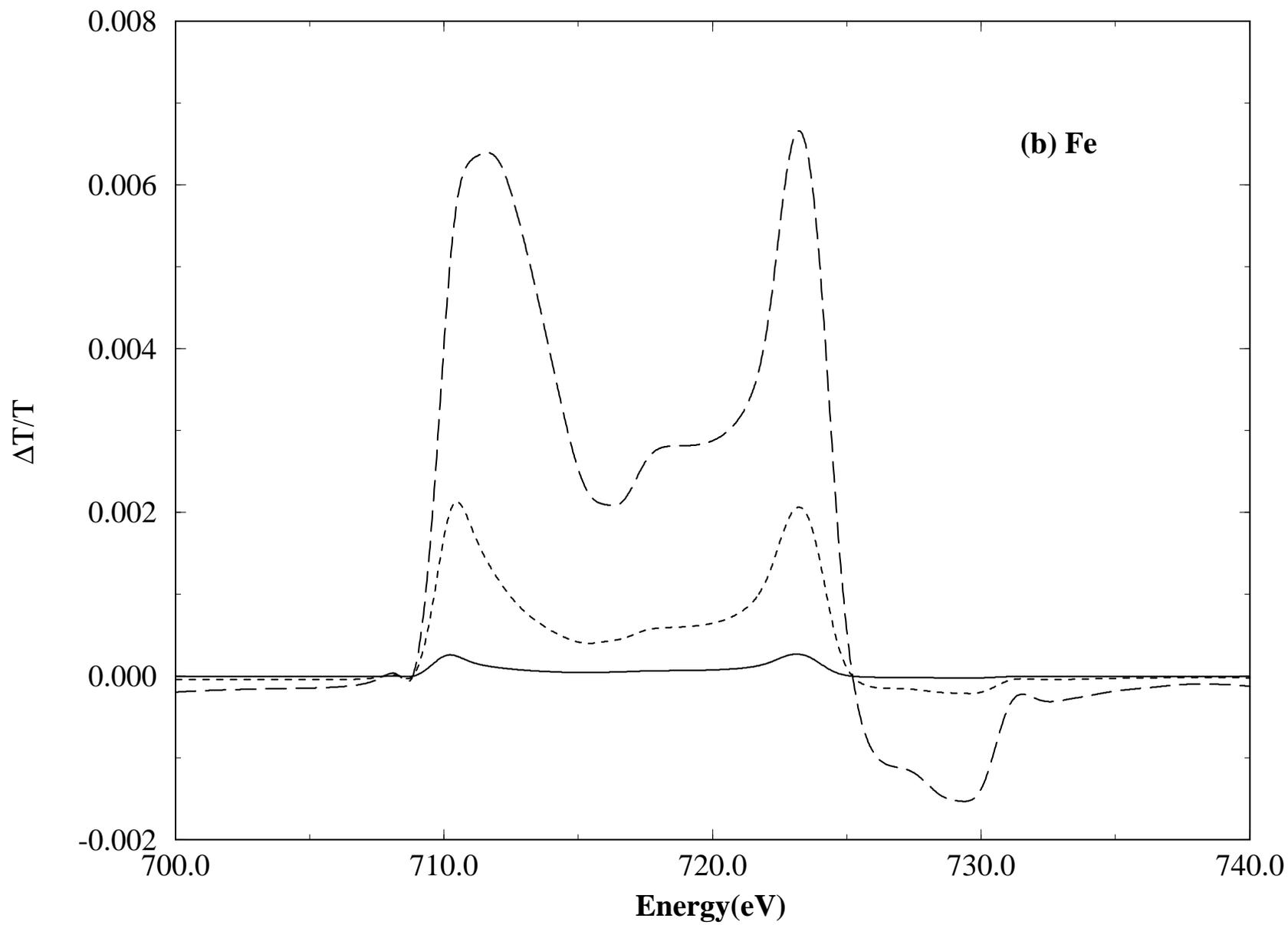
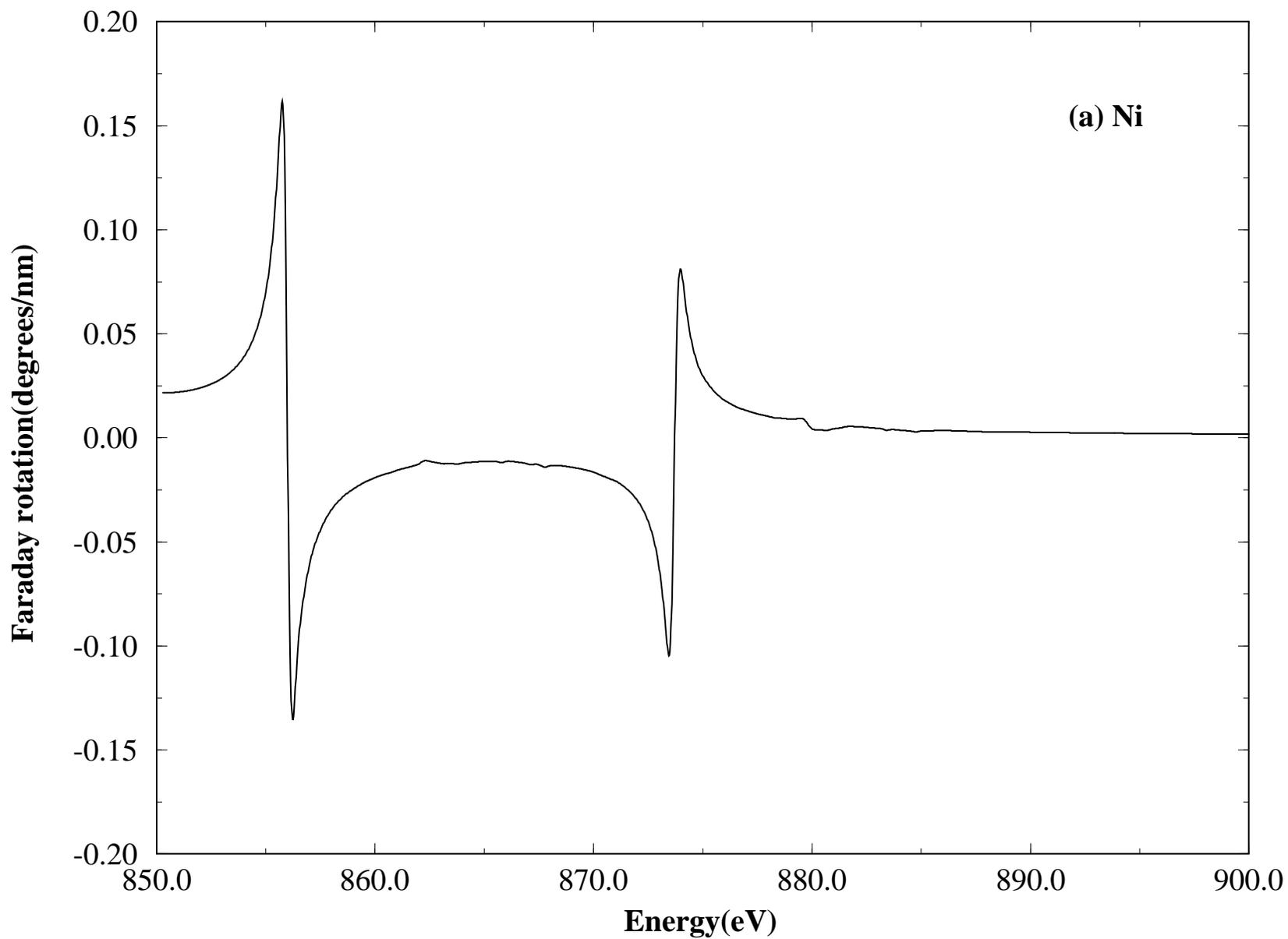
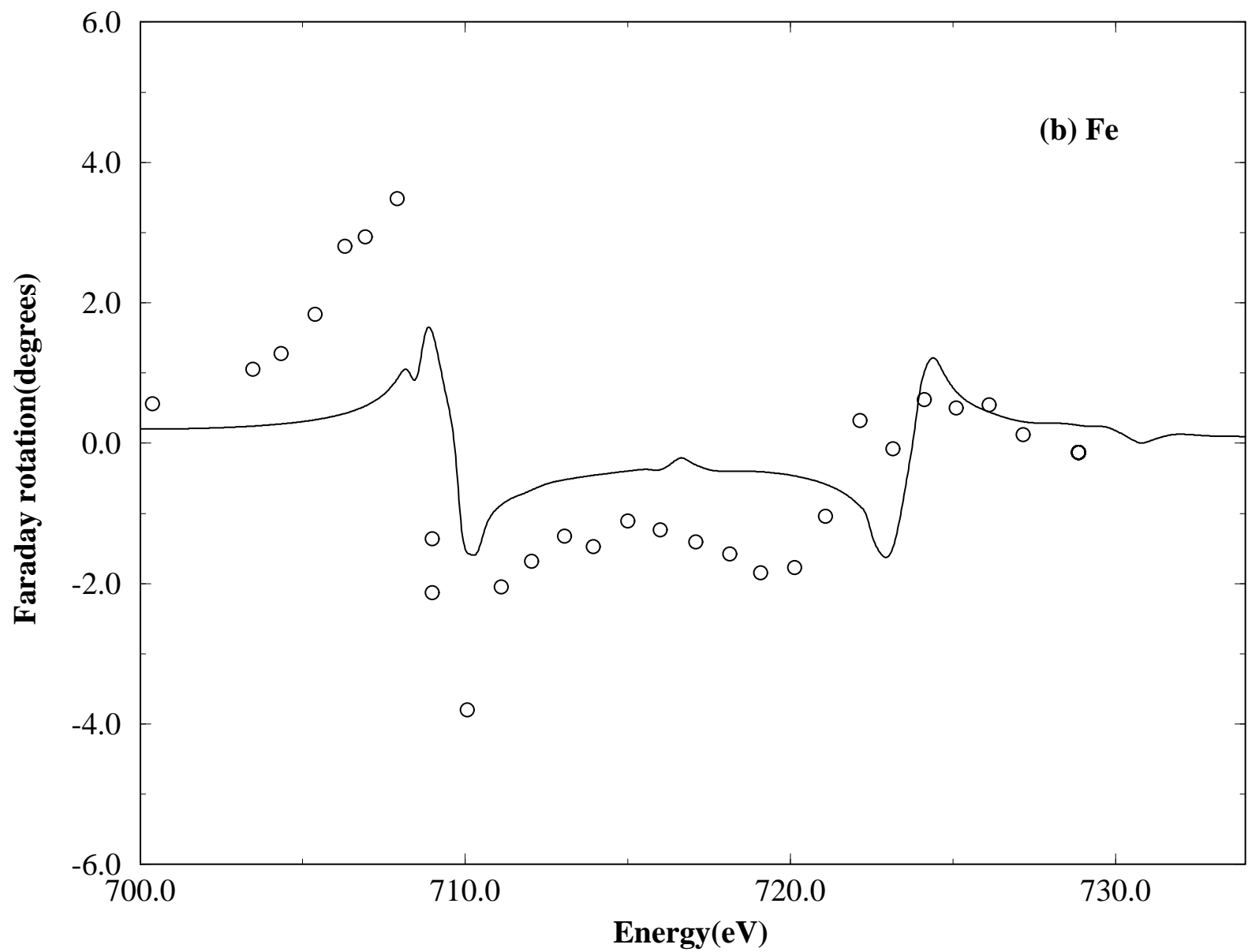
\end{document}